\nonstopmode

\documentclass{JFM-FLM_Au}   
\makeatletter
\def\oddfooterflagdefns#1{\hbox to \hsize{\hfil{\cppagefont\ifx\@volume\undefined\else\textbf{\@volume}\fi\ \ifx\@issue\undefined\ifpaper A\else\ifrapid R\else\ifpersp P\else\iffof F\else X\fi\fi\fi\fi1-\else\ifpaper A\else\ifrapid R\else\ifpersp P\else\iffof F\else X\fi\fi\fi\fi\@issue-\fi\thepage}}}
\def\evenfooterflagdefns#1{\hbox to \hsize{{\cppagefont\ifx\@volume\undefined\else\textbf{\@volume}\fi\ \ifx\@issue\undefined\ifpaper A\else\ifrapid R\else\ifpersp P\else\iffof F\else X\fi\fi\fi\fi1-\else\ifpaper A\else\ifrapid R\else\ifpersp P\else\iffof F\else X\fi\fi\fi\fi\@issue-\fi\thepage}\hfil}}
\makeatother
\usepackage[utf8]{inputenc}
\usepackage{textcomp}
\usepackage{graphicx}
\usepackage{amsmath}
\usepackage{amssymb}
\usepackage{bm}
\usepackage[version=4]{mhchem}
\usepackage{siunitx}
\usepackage{longtable,tabularx}
\usepackage{booktabs}
\usepackage{subcaption}
\usepackage[dvipsnames]{xcolor}
\lefttitle{J. Chen, X. I. A. Yang and P. He}
\righttitle{Journal of Fluid Mechanics}

\title{Automated Discovery of Physically Interpretable Subgrid Length Scales for Detached Eddy Simulation via Agentic AI}

\author{Jiawei Chen\aff{1}, Xiang I.A. Yang\aff{2} \and Ping He\aff{1}}

\affiliation{{\aff{1}Department of Aerospace Engineering, Iowa State University, Ames, Iowa 50011, USA
\aff{2}Department of Mechanical Engineering, Pennsylvania State University, State College, PA, USA}}

\corresau{Ping He, \email{phe@iastate.edu}}

\begin{document}
\maketitle

\begin{abstract}
Traditional turbulence-model development relies heavily on expert knowledge and is time-consuming. This study develops an agentic AI framework, DEScover (automated Discovery of DES closures), to explore subgrid length scales for detached eddy simulation (DES). The user provides only a modeling idea: the length scale is the grid width normal to the local vortex axis. Four large language model (LLM) agents then propose, review, implement, and evaluate candidate formulations and generate scripts for CFD simulations. The Discovery Agent infers the vortex axis from various vortex-identification methods. The candidates are tested on four cases spanning canonical and engineering flows: a flat plate, a backward-facing step, a free shear layer, and an iced airfoil. For all test cases, the best candidate consistently outperforms the standard length scale based on the maximum cell size. Its accuracy is similar to, or slightly better than, that of the expert-designed shear-layer-adapted length scale. Each agentic discovery run takes about 25 minutes on average, excluding the CFD simulations. These results show that LLM-based reasoning can significantly accelerate the discovery of physically interpretable turbulence models and reduce reliance on manual expert formulation.
\end{abstract}

\section{Introduction}
Turbulence modeling is of fundamental importance in computational fluid dynamics (CFD) and has long attracted extensive research attention. Traditionally, the development of turbulence models has relied heavily on both a deep physical understanding of turbulent flows and substantial empirical knowledge. Representative advances include wall-modeling approaches in large eddy simulation (LES), such as integral wall models based on ordinary differential equation (ODE) formulations \citep{yang2015integral}, log-layer mismatch correction and fluctuating wall-stress modeling \citep{yang2017log}, semi-locally scaled eddy-viscosity formulations for high-speed flows \citep{yang2018semi}, and eddy-conductivity closures for wall heat flux prediction in high-speed regimes \citep{chen2022wall}. Additional developments include new length-scale formulations in detached eddy simulation (DES) \citep{zhou2025enhanced}, corrections for large-scale flow separation in Reynolds-averaged Navier–Stokes (RANS) models \citep{chen2024aerodynamic,chen2025separation}, and modifications accounting for rough-wall effects \citep{chen2023roughness}. These advances have significantly improved the predictive capability and applicability of turbulence models for complex flow configurations.

In recent years, the integration of machine learning (ML) and data-driven techniques has emerged as a revolutionary paradigm for turbulence modeling, aiming to overcome the inherent limitations and empirical assumptions of traditional Reynolds-Averaged Navier-Stokes (RANS) models and Large Eddy Simulations (LES). Generally, recent advancements in ML-based turbulence closures can be categorized into five major methodologies. The first relies on data assimilation and field inversion frameworks, which indirectly calibrate empirical coefficients or introduce spatially varying correction source terms into existing transport equations to enhance predictive accuracy \citep{rincon2025generalisable,bin2023constrained,bin2024constrained,hu2025data}. Notably, recent strides have successfully extended these frameworks from steady states to time-accurate unsteady flows and automated aerodynamic shape optimization \citep{fang2024field,li2025field}. The second approach involves direct corrections targeting the anisotropy tensor of Reynolds stresses, where novel architectures like Tensor Basis Neural Networks (TBNNs) are utilized to explicitly preserve Galilean invariance and physical symmetries \citep{ling2016reynolds}. Subsequent extensions successfully resolved the numerical instability of embedding these learned structures back into CFD solvers \citep{wu2018physics}, and further generalized the invariant tensor architectures to turbulent scalar and heat flux modeling \citep{milani2021turbulent}. To address the ``black-box'' nature and lack of interpretability of deep neural networks, the third category leverages symbolic regression and gene expression programming to discover explicit, algebraic turbulence formulations directly from high-fidelity data \citep{yang2025data}. Notably, this data-driven technique has been successfully utilized to discover explicit correction functions for the baseline shear-stress-transport (SST) model, significantly balancing the interpretability and generalization capability of RANS closures \citep{wu2023enhancing}. In the fourth category, Physics-Informed Machine Learning (PIML) \citep{raissi2018physicsinformed} has gained substantial traction, wherein fundamental conservation laws and boundary conditions are embedded into the network via customized loss functions or hard-constrained architectures \citep{yang2019predictive}. This paradigm has been robustly applied to derive data-driven subgrid-scale closures for LES via physical deconvolution frameworks \citep{maulik2018data} as well as solving unclosed RANS systems directly via automatic differentiation of partial differential equation residuals \citep{eivazi2022physics}. Beyond these predominantly supervised paradigms, the fifth category leverages Deep Reinforcement Learning (DRL) to resolve the ``\textit{a priori} vs.\ \textit{a posteriori}'' discrepancy by framing the CFD solver as the environment. This methodology has been pioneered to automatically discover dynamic subgrid-scale closures for LES \citep{novati2021automating}, construct generalized near-wall models \citep{bae2022scientific}, and recently advanced into high-performance computing (HPC) platforms through a solver-agnostic, open-source SmartFlow framework \citep{xiao2025smartflow}. Despite this progress, data-driven turbulence models still have limitations. First, most of them need a large amount of high-fidelity data for training. Such data are expensive to obtain for complex flows with high Reynolds numbers. Second, neural-network-based turbulence model augmentations are often black boxes, making their predictions difficult to interpret physically and limiting their portability across CFD solvers. Although symbolic regression and gene expression programming can produce explicit mathematical expressions based on flow features, the mathematical relationships among these features are primarily determined through data-driven fitting rather than physical reasoning. Consequently, the resulting formulations may still lack clear physical connections to established turbulence theories.

The rapid development of large language models (LLMs), trained on vast
amounts of source code and scientific literature, has opened a new paradigm for automated
scientific discovery. Early studies mainly focused on reducing the operational barrier of CFD by allowing users to describe simulation tasks in natural language. Representative systems, such as OpenFOAMGPT \citep{pandey2025openfoamgpt}, OpenFOAMGPT 2.0 \citep{feng2026openfoamgpt}, Foam-Agent/Foam-Agent 2.0 \citep{pan2025foamagent2,pan2026automatingcfd}, FoamGPT \citep{pan2025foamgpt}, CFDagent \citep{xu2025cfdagent}, ChatCFD \citep{fan2025chatcfd}, FlamePilot \citep{xiao2026combustion}, and natural-language-to-simulation intermediate-representation frameworks \citep{shenoy2026natural}, demonstrate that LLM-based agents can automate case setup, mesh generation, solver configuration, error correction, post-processing, and visualization across OpenFOAM-based and other CFD environments. A related line of work extends this capability to aerodynamic and multidisciplinary design optimization, where multi-agent systems such as specialized aerodynamic design agents \citep{lee2025aerodynamic}, AirfoilAgent \citep{fan2026airfoilagent}, iDesign \citep{idesign2026conversational}, MDO agents \citep{zhang2025mdoagent}, vision-language models for engineering design evaluation \citep{ahmed2024concept}, AI design agents for aesthetic and aerodynamic car design \citep{elrefaie2025ai}, and risk-aware set-based engineering design agents \citep{karniadakis2026risk} couple LLMs with CAD, CFD, optimization, and knowledge-retrieval tools to support geometry generation, design-space exploration, and iterative performance improvement. A deeper and more impactful application lies in employing LLMs to propose, refine, and
discover new turbulence modelling paradigms---a direction that has only recently begun to
be explored.
Two complementary paradigms have recently emerged for LLM-assisted turbulence modelling.
\citet{liu2025autoturb} embedded LLMs within an evolutionary search framework to
automatically discover algebraic Reynolds-stress corrections, treating the LLM as an
expression generator guided by a quantitative fitness function.
\citet{yang2025turbulence} adopted a contrasting approach, engaging DeepSeek-R1 as a
reasoning collaborator through iterative human--AI dialogue, successfully developing
near-wall models for LES under adverse pressure gradients, system rotation, and surface
roughness.
These works demonstrate that LLMs can contribute to turbulence modelling as physics-aware
reasoning agents. 


In this paper, a multi-agent framework, termed DEScover (automated Discovery of DES closures), is proposed to systematically explore candidate subgrid length scales for detached eddy simulations (DES). 
The primary contribution of this work is the automation of the turbulence model development process, from an initial modeling idea to explicit mathematical formulations, code implementation, and \textit{a posteriori} CFD evaluations across multiple flow cases. Compared with existing data-driven turbulence model augmentation methods, the proposed agentic-AI-driven approach generates explicit formulations with clearer connections to underlying turbulence physics. Four flow cases are considered for validation: a flat plate, a backward-facing step, a free shear layer, and an iced airfoil. The flat-plate case assesses whether the candidate length scales preserve accurate predictions for attached flow. The backward-facing step and free-shear-layer cases evaluate their ability to alleviate the delayed onset of Kelvin--Helmholtz (K--H) instability. The iced-airfoil case represents a high-Reynolds-number engineering flow and provides a more challenging test of the generalization capability of the candidate length scales. 

The remainder of this paper is organized as follows. \S\,\ref{sec:methodology} introduces the methodology. \S\,\ref{sec:results} presents the results and discussion. Finally, \S\,\ref{sec:conclusion} provides the conclusions of this study.

\section{Methodology}
\label{sec:methodology}

\subsection{IDDES Method}
\label{sec:iddes}
\subsubsection{IDDES equation}
The improved delayed detached eddy simulation (IDDES) method employed in this study is based on the one-equation Spalart-Allmaras (SA) turbulence model. In this approach, the transport equation for the working variable $\tilde{\nu}$ is modified by replacing the distance to the wall $d$ in the destruction term with a hybrid IDDES length scale $l_{IDDES}$ \citep{shur2008hybrid}. The transport equation for $\tilde{\nu}$ is expressed as:

\begin{equation}
  \frac{\partial(\rho \tilde{\nu})}{\partial t} 
  + \nabla\cdot(\rho \mathbf{u} \tilde{\nu}) 
  = \frac{1}{\sigma} \nabla\cdot \left[ (\mu + \rho\tilde{\nu}) \nabla \tilde{\nu} \right] 
  + \frac{C_{b2}\rho}{\sigma} |\nabla \tilde{\nu}|^2 
  + \rho P_{\tilde{\nu}} 
  - \rho C_{w1} f_w \left( \frac{\tilde{\nu}}{l_{\text{IDDES}}} \right)^2
  \label{eq:nu-transport}
\end{equation}

where $\tilde{\nu}$ is the modified turbulent kinematic viscosity, $\mathbf{u}$ is the velocity, and $P_{\tilde{\nu}}$ represents the production term. The IDDES length scale $l_{IDDES}$ is defined to provide a seamless transition between RANS and LES modes:

\begin{equation}
  l_{\text{IDDES}} = \tilde{f}_d (1 + f_e) d + (1 - \tilde{f}_d) l_{\text{LES}}
  \label{eq:l_iddes}
\end{equation}

In this expression, $d$ represents the RANS length scale (distance to the closest wall), and $l_{LES}$ is the LES subgrid length scale. The empirical blending function $\tilde{f}_d$ is designed to prevent Grid-Induced Separation (GIS) and to provide protection against log-layer mismatch. The LES subgrid length scale is defined as $l_{\text{LES}} = C_{\text{DES}} \Psi \Delta$, where $C_{DES}$ is a model constant, $\Psi$ is a low-Reynolds number correction, and $\Delta$ is the grid filter width. In the present implementation within OpenFOAM 2312, the grid filter width $\Delta$ follows the specialized IDDES formulation to handle highly anisotropic meshes. According to the IDDESDelta implementation, $\Delta$ is determined by:

\begin{equation}
\Delta = \min \left[ \max \left( C_w d_w, C_w \Delta_{\text{max}}, \Delta_{f2f} \right), \Delta_{\text{outer}} \right]
    \label{eq:iddes_delta}
\end{equation}

where $d_w$ is the wall-normal distance, $C_w$ is a model constant (typically 0.15), and $\Delta_{f2f}$ represents the maximum face-to-face distance along the wall-normal direction. The term $\Delta_{\text{max}}$ represents the maximum local cell edge length. Crucially, $\Delta_{\text{outer}}$ serves as the upper-bound clip for the grid filter width, which plays a vital role in determining the transition from RANS to LES in the outer regions of the boundary layer and in free shear layers. In the standard IDDES, $\Delta_{\text{outer}}=\Delta_{\max}$.

Furthermore, in
RANS-modeled attached boundary layers, the LES length scale may
become smaller than the RANS length scale, potentially contaminating
RANS functionality and resulting in lower wall friction values. To mitigate these issues, we restrict the LES length scale as follows

\begin{equation}
\Delta_{\text{outer},lim} =
\begin{cases}
\max \left( \Delta_{\text{outer}}, C_{lim} V^{1/3} \right), & f_d < \varepsilon \\
\Delta_{\max}, & f_d \ge \varepsilon,
\end{cases}
\label{eq:cdes_amd_lim}
\end{equation}

where $V$ is the cell volume, and $C_{lim}$ and $\varepsilon$ are constants, set to 0.15 and 0.01, respectively. The LES length scale then becomes

\begin{equation}
l_{\text{LES}} = C_{\text{DES}}\,\Psi\min \left[ \max \left( C_w d_w, C_w \Delta_{\text{max}}, \Delta_{f2f} \right), \Delta_{\text{outer},lim} \right]
    \label{eq:l_les}
\end{equation}

The standard $\Delta_{\max} = \max(\Delta x,
\Delta y, \Delta z)$
definition is known to be sensitive to highly anisotropic cells (e.g., pencil cells with
$\Delta x \sim \Delta y \ll \Delta z$, and book cells with $\Delta y \ll \Delta x \sim
\Delta z$), where it can artificially delay the RANS-to-LES transition and broaden the
grey area. The present work focuses on the discovery of improved expressions for $\Delta_{\text{outer}}$ to enhance the model's robustness and accuracy under complex anisotropic grid conditions, where the default definition may lead to an artificial delay of the RANS-to-LES transition.

\begin{figure}[hbt!]
  \centering
  \begin{subfigure}[b]{0.26\textwidth}
    \includegraphics[width=\textwidth]{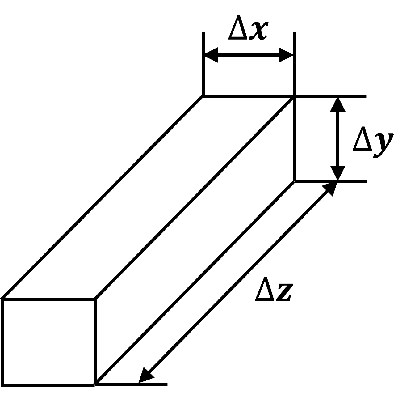}
    \caption{Pencil cell}
  \end{subfigure}
  \hfill
  \begin{subfigure}[b]{0.40\textwidth}
    \includegraphics[width=\textwidth]{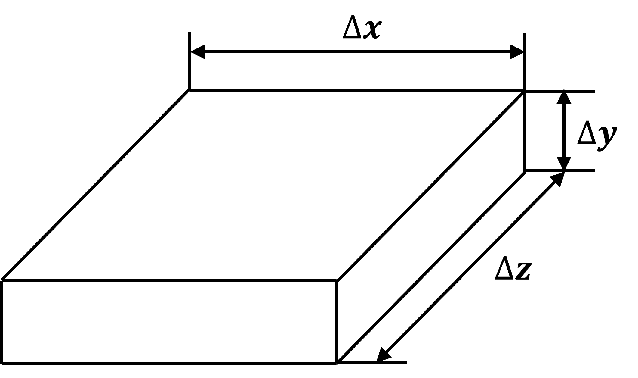}
    \caption{Book cell}
  \end{subfigure}
  \caption{Two types of anisotropic grids encountered in engineering applications: (a) pencil cell, characterized by $\Delta x \approx \Delta y \ll \Delta z$, and (b) book cell, characterized by $\Delta y \ll \Delta x \approx \Delta z$.}
  \label{fig:anisotropic_cells}
\end{figure}

\subsubsection{Length scales}

Conventional length scales employed in subgrid-scale (SGS) models, such as the maximum cell dimension $\Delta_{max}$ (\eqref{eq:delta_max}) and the cube root of cell volume $\Delta_{vol}$ (\eqref{eq:delta_vol}), rely exclusively on local grid metrics without consideration of flow characteristics:

\begin{equation}
\Delta_{\max} = \max(\Delta_x, \Delta_y, \Delta_z)
\label{eq:delta_max}
\end{equation}

\begin{equation}
\Delta_{vol} = \sqrt[3]{\Delta_x \Delta_y \Delta_z}
\label{eq:delta_vol}
\end{equation}

where $\Delta_x$, $\Delta_y$ and $\Delta_z$ represent the cell spacing in the x, y, and z directions, respectively.

\begin{figure}[hbt!]
  \centering
  \includegraphics[width=0.3\textwidth]{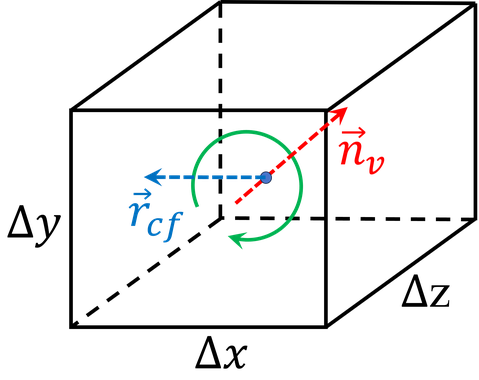}
  \caption{Geometric illustration of the vortex-axis-based length scale $\Delta_{\perp}$.}
  \label{fig:rotation_axis}
\end{figure}

In contrast, the modeling concept considered in this study is a flow-based length scale that adapts to local turbulent structures. Recognizing that turbulent dissipation occurs predominantly in directions perpendicular to the rotation axis, we define the length scale as the characteristic dimension perpendicular to the local vortex orientation. Figure~\ref{fig:rotation_axis} shows the geometric illustration of the vortex-axis-based length scale $\Delta_{\perp}$. Equation~\eqref{eq:delta_perp} quantifies twice the maximum magnitude of the cross product between the unit rotation vector and the cell center-to-face vectors. Geometrically, this operation identifies the maximum characteristic length perpendicular to the vortical axis within each computational cell. This approach fundamentally differs from conventional grid-based length scales by directly accounting for the orientation of the local vortical structures.

This length scale, denoted as $\Delta_{\perp}$, is computed using:

\begin{equation}
\Delta_{\perp} = C_\delta \max_f (2 | \mathbf{n}_v \times \mathbf{r}_{f} |), \quad f=1,\ldots,6
\label{eq:delta_perp}
\end{equation}

Equation~\eqref{eq:delta_perp}, including the coefficient $C_\delta$, is the modeling concept prescribed by the user in the prompt of the DEScover workflow (figure~\ref{fig:prompt}). The method for determining the vortex-axis direction $\mathbf{n}_v$ is not prescribed and is left to the Discovery Agent. Consequently, the candidate length scales $\Delta_{\omega}$, $\Delta_{\text{swirl}}$, $\Delta_{R}$, and $\Delta_{\lambda_2}$ examined in this study all take the form of equation~\eqref{eq:delta_perp} and differ only in how $\mathbf{n}_v$ is determined (\S\,\ref{sec:candidates} and Appendix~\ref{app:proposals}).

The coefficient $C_\delta = 1.035$ in the formulation was derived as a statistical calibration constant. $\mathbf{n}_{v}$ is the unit vector in the vortical axis direction, and $\boldsymbol{r}_f$ is the vector from the cell centroid to the centroid of face $f$. On a uniform isotropic Cartesian grid with $\Delta x = \Delta y = \Delta z = h$, the unscaled maximum reduces to

\begin{equation}
\max_{f} \left( 2 \left| \mathbf{n}_v \times \boldsymbol{r}_f \right| \right) = h \max\left(\sqrt{1-a^2},\, \sqrt{1-b^2},\, \sqrt{1-c^2}\right),
\end{equation}

where $(a, b, c)$ are the Cartesian components of $\mathbf{n}_v$. Because the vortical direction varies across the flow field, we required that the ensemble average of $\Delta_{\perp}$ over all orientations of $\mathbf{n}_v$, uniformly distributed on the unit sphere, recovers the standard grid spacing $h$. Evaluating this average yields

\begin{equation}
\mathbb{E}\left[\max\left(\sqrt{1-a^2},\, \sqrt{1-b^2},\, \sqrt{1-c^2}\right)\right] \approx 0.9666,
\end{equation}

so that the calibration coefficient is

\begin{equation}
C_\delta = \frac{1}{0.9666} \approx 1.035.
\end{equation}

Geometrically, this correction accounts for the fact that, on average over all vorticity orientations, the cross-product-based projection underestimates the true cell width by approximately $3.3\%$; multiplying by $C_\delta$ restores statistical consistency with the nominal grid spacing on isotropic meshes.

\begin{figure}[hbt!]
  \centering
  \begin{subfigure}[b]{0.45\textwidth}
    \includegraphics[width=\textwidth]{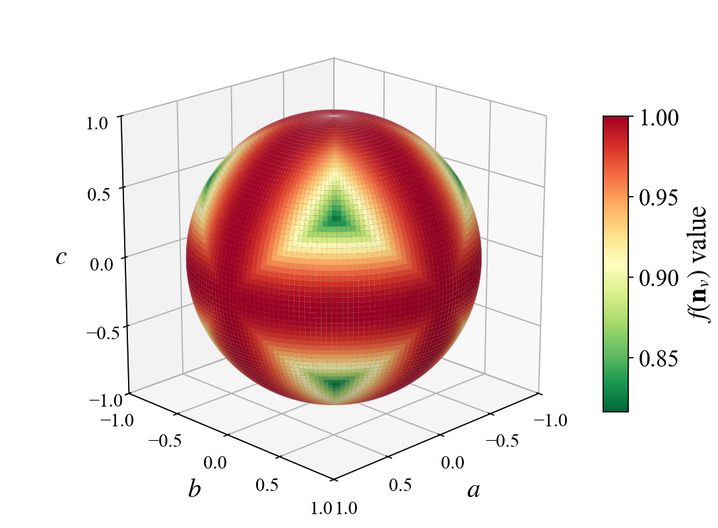}
    \caption{Distribution of $f(\bm{n}_v)$ on the unit sphere}
  \end{subfigure}\hfill
  \begin{subfigure}[b]{0.45\textwidth}
    \includegraphics[width=\textwidth]{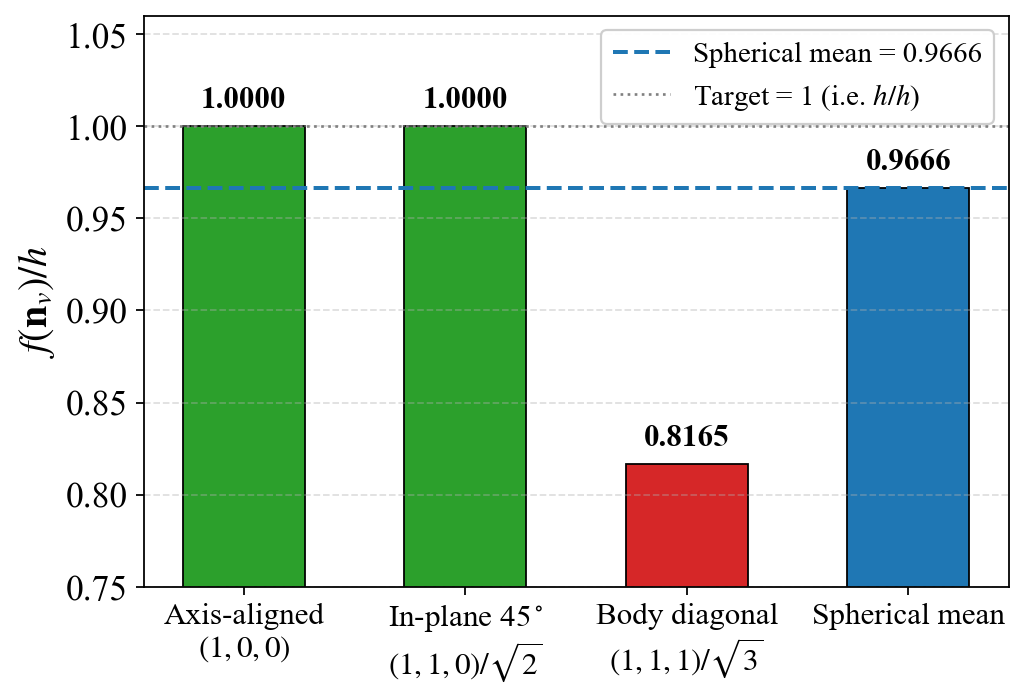}
    \caption{$f(n_v)$ for three representative vortex orientations}
  \end{subfigure}
  \begin{subfigure}[c]{0.45\textwidth}
    \includegraphics[width=\textwidth]{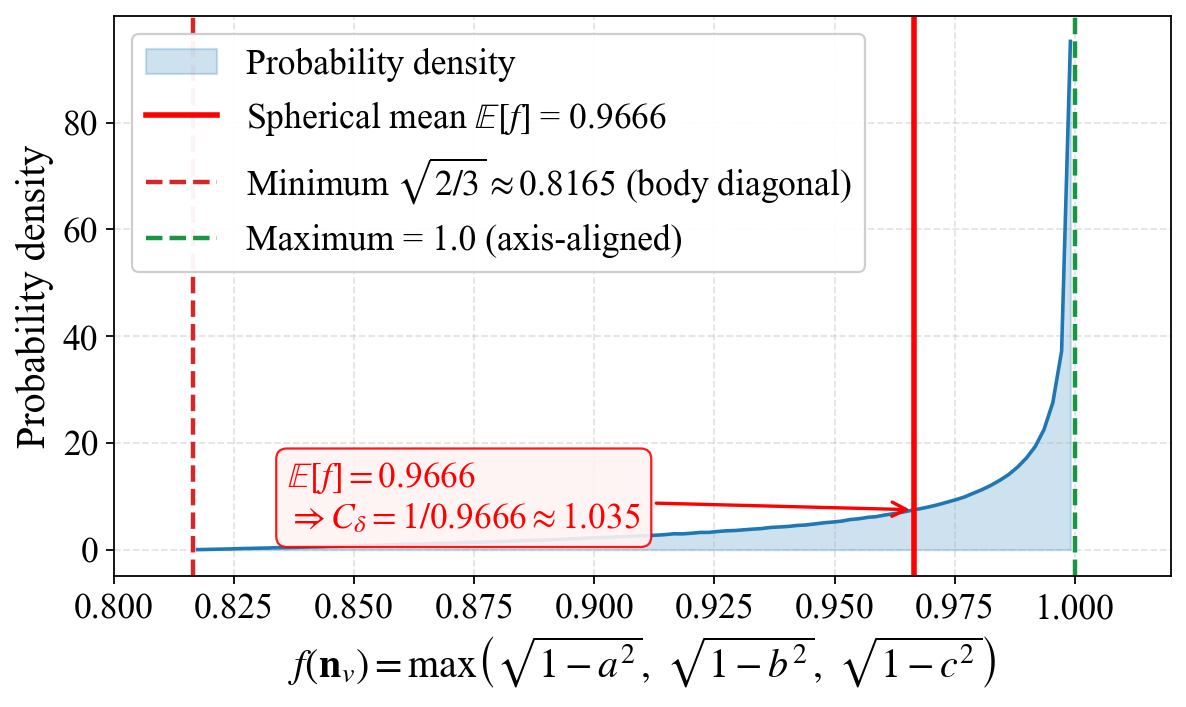}
    \caption{Probability density of $f(n_v)$}
  \end{subfigure}
  \caption{Geometric interpretation and statistical derivation of the correction coefficient $C_\delta = 1.035$ in the vortex-aware filter-width formula.}
  \label{fig:C_delta}
\end{figure}

Figure~\ref{fig:C_delta} shows the geometric interpretation and statistical derivation of the correction coefficient $C_\delta$. The function $f(n_v) = \max\left(\sqrt{1-a^2},\,\sqrt{1-b^2},\,\sqrt{1-c^2}\right)$, where $(a,b,c)$ denote the direction cosines of the vortical axis $n_v$, governs the sensitivity of the proposed filter width to the local vortex orientation. As shown in figure~\ref{fig:C_delta}(a), $f(n_v)$ varies continuously over the unit sphere: it reaches its maximum value of unity whenever the vortex is aligned with a coordinate axis or lies in a coordinate plane (i.e., at least one direction cosine vanishes), and attains its global minimum of $\sqrt{2/3}\approx 0.8165$ when all three components are equal, corresponding to the body-diagonal direction. This directional anisotropy reflects the fundamental geometric relationship between the vortex axis and the faces of a hexahedral mesh cell. Figure~\ref{fig:C_delta}(b) further quantifies $f(n_v)$ for three representative vortex orientations on an isotropic mesh ($h_x=h_y=h_z=h$): the axis-aligned case $(1,0,0)$ and the in-plane diagonal $(1,1,0)/\sqrt{2}$ both yield $f=1$, whereas the body-diagonal direction $(1,1,1)/\sqrt{3}$ gives the minimum value of $0.8165$. To remove this directional bias, $C_\delta$ is determined by requiring that the spherical mean of $f(n_v)$ equal unity. Figure~\ref{fig:C_delta}(c) presents the probability density of $f(n_v)$ evaluated from $2\times10^6$ Monte Carlo samples drawn uniformly over the unit sphere, revealing a strongly right-skewed distribution concentrated near the maximum. The resulting spherical mean is $\mathbb{E}[f]\approx 0.9666$, so that the normalisation condition $C_\delta\cdot\mathbb{E}[f]=1$ gives $C_\delta = 1/0.9666\approx 1.035$. This value ensures that, on a statistically isotropic distribution of vortex directions, the expected filter width returned by the proposed formula equals the local mesh size $h$, thereby preserving consistency with conventional LES filter-width definitions. This value is also consistent with the default coefficient (1.035) of the face-centre-based vorticity-direction length scale $\tilde{\Delta}_{\omega}$ in the OpenFOAM implementation of the SLA length scale (see the description following equation~\eqref{eq:delta_sla}), which provides an independent confirmation of the present calibration.

In addition to $\Delta_{\max}$, the shear-layer-adapted (SLA) length scale of \citet{shur2015enhanced} is included as a second baseline. $\Delta_{\text{SLA}}$ is an existing, expert-designed flow-dependent length scale that was developed specifically to accelerate the RANS-to-LES transition in separating shear layers. It is defined as
\begin{equation}
\Delta_{\text{SLA}} = \tilde{\Delta}_{\omega}\, F_{KH}\left(\langle VTM \rangle\right),
\label{eq:delta_sla}
\end{equation}
where $\tilde{\Delta}_{\omega}$ is the vorticity-direction-based length scale first published by \citet{mockett2015two}, which builds on the idea of \citet{chauvet2007zonal} of accounting for the direction of the vorticity vector on strongly anisotropic cells. For a hexahedral cell with vertices located at $\mathbf{r}_n$ ($n=1,\ldots,8$), it reads
\begin{equation}
\tilde{\Delta}_{\omega} = \frac{1}{\sqrt{3}} \max_{n,m=1,\ldots,8} \left| \mathbf{l}_n - \mathbf{l}_m \right|, \qquad \mathbf{l}_n = \mathbf{n}_{\omega} \times \mathbf{r}_n,
\label{eq:delta_omega_tilde}
\end{equation}
in which $\mathbf{n}_{\omega}$ is the unit vector aligned with the vorticity vector. $\tilde{\Delta}_{\omega}$ is the diameter of the set of cross-product points $\mathbf{l}_n$ divided by $\sqrt{3}$, which approximately recovers $\Delta_{\max}$ for three-dimensional turbulence computed on cubic cells. Because $\tilde{\Delta}_{\omega}$ alone is insufficient to unlock the K--H instability when the grid is marginal in resolving it, \citet{shur2015enhanced} further introduced the function $F_{KH}$ of the vortex tilting measure ($VTM$), a kinematic sensor that identifies the quasi-two-dimensional flow regions typical of the initial region of shear layers:
\begin{equation}
VTM = \frac{\sqrt{6}\,\left| (\hat{\mathbf{S}}\cdot\boldsymbol{\omega})\times\boldsymbol{\omega} \right|}{\omega^2 \sqrt{3\,\mathrm{tr}(\hat{\mathbf{S}}^2) - \left[\mathrm{tr}(\hat{\mathbf{S}})\right]^2}}\,\max\left\{1, \frac{\nu^*}{\nu_t}\right\}, \qquad \nu^* = 0.2\,\nu,
\label{eq:vtm}
\end{equation}
\begin{equation}
F_{KH} = \max\left\{ F_{KH}^{\min}, \min\left\{ F_{KH}^{\max},\; F_{KH}^{\min} + \frac{F_{KH}^{\max}-F_{KH}^{\min}}{a_2-a_1}\left(\langle VTM\rangle - a_1\right) \right\} \right\},
\label{eq:fkh}
\end{equation}
where $\hat{\mathbf{S}}$ is the strain tensor, $\nu$ and $\nu_t$ are the molecular and eddy kinematic viscosities, and $\langle VTM\rangle$ denotes the average of $VTM$ over the current and closest neighboring cells. The factor $\sqrt{6}$ bounds $VTM$ between 0 and 1: $VTM$ is close to zero in quasi-two-dimensional regions, where the vorticity is aligned with an eigenvector of the strain tensor, and close to unity in developed three-dimensional turbulence. The factor $\max\{1,\nu^*/\nu_t\}$ deactivates $F_{KH}$ in the inviscid region, where $VTM$ oscillates strongly because the vorticity is close to zero. The empirical constants are $F_{KH}^{\min}=0.1$, $F_{KH}^{\max}=1.0$, $a_1=0.15$, and $a_2=0.3$ \citep{shur2015enhanced}. To preserve the shielding of attached boundary layers in DDES, \citet{shur2015enhanced} further replaced $F_{KH}$ with
\begin{equation}
F_{KH}^{lim} =
\begin{cases}
1.0, & f_d < 1-\varepsilon_{\text{SLA}}, \\
F_{KH}, & f_d \ge 1-\varepsilon_{\text{SLA}},
\end{cases}
\qquad \varepsilon_{\text{SLA}} = 0.01,
\label{eq:fkh_lim}
\end{equation}
where $f_d$ is the DDES delay function \citep{shur2015enhanced}. In the OpenFOAM implementation (\texttt{SLADelta}) used in this study, the shielding of equation~\eqref{eq:fkh_lim} is activated with $\varepsilon_{\text{SLA}}=0.01$, and $f_d$ is evaluated within the length-scale class as
\begin{equation}
f_d = 1 - \tanh\left[\left(C_{d1}\, r_d\right)^{C_{d2}}\right], \qquad r_d = \min\left\{\frac{\nu_t + \nu}{|\nabla \mathbf{U}|\,\kappa^2 d_w^2},\, 10\right\},
\label{eq:fd_sla}
\end{equation}
with $\kappa=0.41$, $C_{d1}=20$, and $C_{d2}=3$. In addition, the implementation evaluates $\tilde{\Delta}_{\omega}$ in the face-centre-based form $1.035\,\max_f\left(2\left|\mathbf{n}_{\omega}\times\mathbf{r}_f\right|\right)$ instead of the vertex-based form of equation~\eqref{eq:delta_omega_tilde}. This form is identical to the candidate $\Delta_{\omega}$ of Proposal~1 (Appendix~\ref{app:proposals}). Consequently, in the present simulations $\Delta_{\text{SLA}} = \Delta_{\omega}\,F_{KH}^{lim}$, and the comparison between $\Delta_{\omega}$ and $\Delta_{\text{SLA}}$ isolates the effect of the kinematic sensor $F_{KH}^{lim}$. Compared with equation~\eqref{eq:delta_perp}, $\Delta_{\text{SLA}}$ achieves the reduction of the length scale in the initial region of the shear layer through two ingredients: the vorticity-direction-based $\tilde{\Delta}_{\omega}$ and the additional kinematic sensor $F_{KH}$. In contrast, equation~\eqref{eq:delta_perp} contains no additional sensor, and the reduction relative to $\Delta_{\max}$ results from the vortex-axis direction $\mathbf{n}_v$ alone.

\subsection{The DEScover Framework}
\label{sec:autodes}

\subsubsection{Overall Architecture}
\label{sec:architecture}

\begin{figure}[hbt!]
  \centering
  \includegraphics[width=1.0\textwidth]{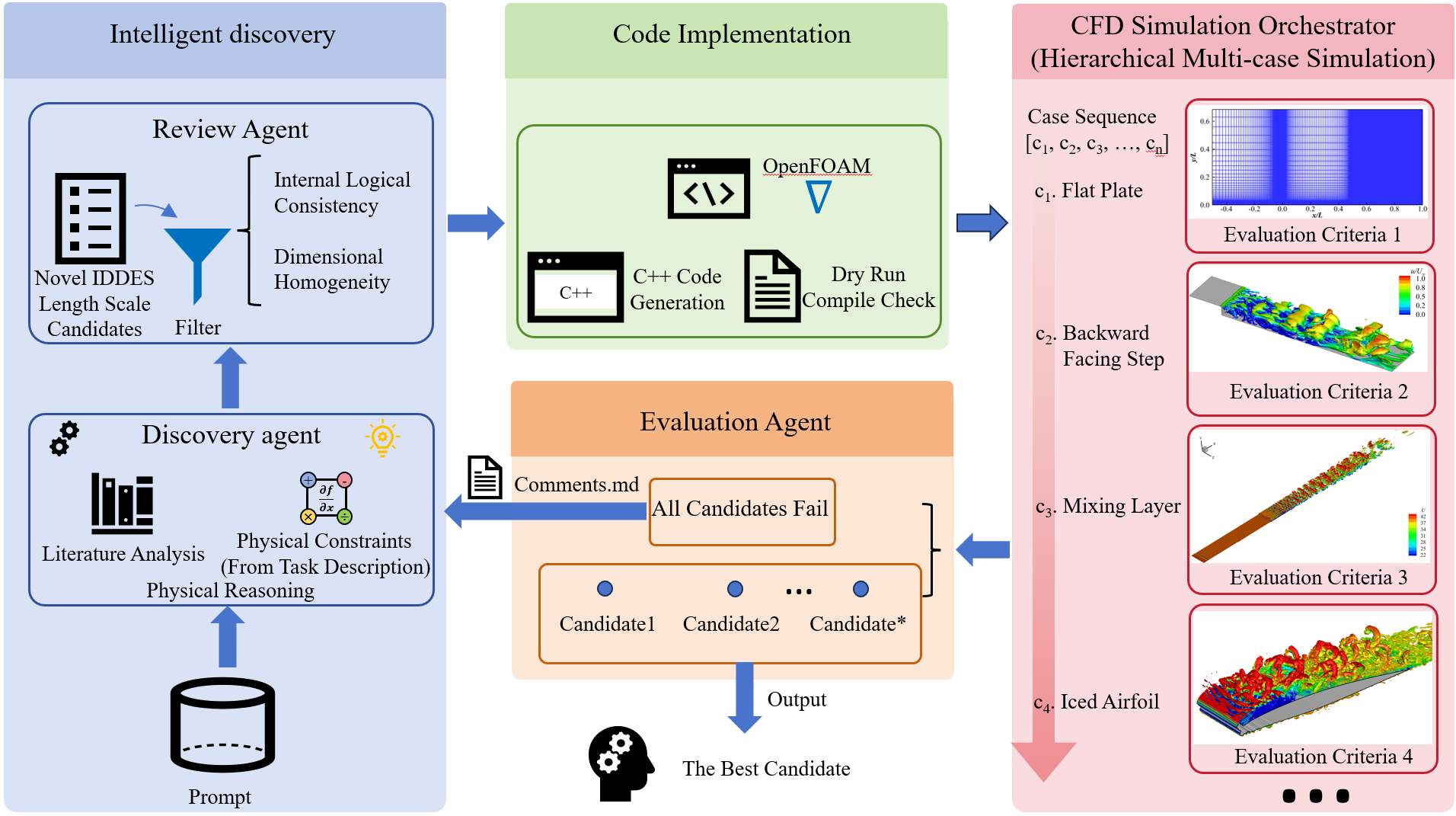}
  \caption{Schematic of the multi-case, closed-loop DEScover (automated Discovery of DES closures) framework.}
  \label{fig:AutoTM}
\end{figure}

This section presents an automated framework for turbulence modeling. The proposed framework is multi-case and operates in a closed-loop manner. DEScover targets the model closures of DES. In the present study, it is applied to the subgrid length scale of IDDES. The workflow itself does not depend on a specific turbulence model. Its extension to RANS and LES models is left for future work. The overall architecture consists of several tightly coupled modules, as illustrated in
figure~\ref{fig:AutoTM}. The framework consists of four LLM-based agents and one script-based module. The four agents are the Discovery Agent, the Review Agent, the Code Implementation Agent, and the Evaluation Agent. The fifth module is the CFD Simulation Orchestrator. It is a Python script and does not use an LLM. It runs the CFD simulations of the case sequence and collects the results. All modules are automatically invoked through Python scripts to ensure the reliable operation of the framework. The overall workflow is as follows. First, the user defines a prompt and submits it to the Discovery Agent. The Discovery Agent then generates candidate models according to the user-defined prompt. Next, the Review Agent examines the candidate models to ensure their logical consistency and dimensional correctness. The reviewed models are then transferred to the Code Implementation Agent. After code generation, the CFD Simulation Orchestrator applies the candidate models to multiple benchmark cases. To reduce the total execution time of the CFD Simulation Orchestrator, each candidate model is assigned to an independent but identical case sequence, $[c_1, c_2, c_3, ..., c_n]$, and all sequences are executed in parallel. When all CFD simulations are completed, the Evaluation Agent analyzes the simulation results and determines whether a satisfactory model has been obtained. If a candidate model meets all the predefined evaluation criteria, it is selected as the final output. Otherwise, feedback is generated and returned to the Discovery Agent to refine the original user-defined prompt. This iterative process continues until a satisfactory model is obtained.

The agents do not rely only on the general knowledge of the LLM. Each agent follows a task-specific instruction file, called Agent Skill. The skill bounds the behavior of the agent and makes its output more robust and repeatable. Each skill defines four items: the input files, a step-by-step procedure, the rules that must be followed, and a fixed output format. For the Discovery Agent, the input is the user prompt (figure~\ref{fig:prompt}). Each alternative method must be written as a separate proposal. Each proposal must follow a fixed template. The template contains the full expression, a table of variable definitions, a step-by-step explanation, the applicability conditions, and the advantages and limitations. For the Review Agent, the instructions are the four review criteria given in \S\,\ref{sec:discovery}. For the Code Implementation Agent, the instructions require the agent to first search the OpenFOAM source code for an existing equivalent method. An existing method is not implemented again. Instead, it is used as a template for the coding style and the class structure. The agent must write the code only into a specified directory. It must also add a header that links the code to its proposal file and to the template file. The agent then checks whether the code is consistent with the formula (\S\,\ref{sec:openfoam}). For the Evaluation Agent, the instructions specify the evaluation criteria of each case and the format of the feedback file \texttt{comments.md} (\S\,\ref{sec:closed_loop}). The CFD Simulation Orchestrator does not use an LLM. Its behavior is fully fixed by the Python script.

\subsubsection{Intelligent discovery}
\label{sec:discovery}
\begin{figure}[hbt!]
  \centering
  \includegraphics[width=0.9\textwidth]{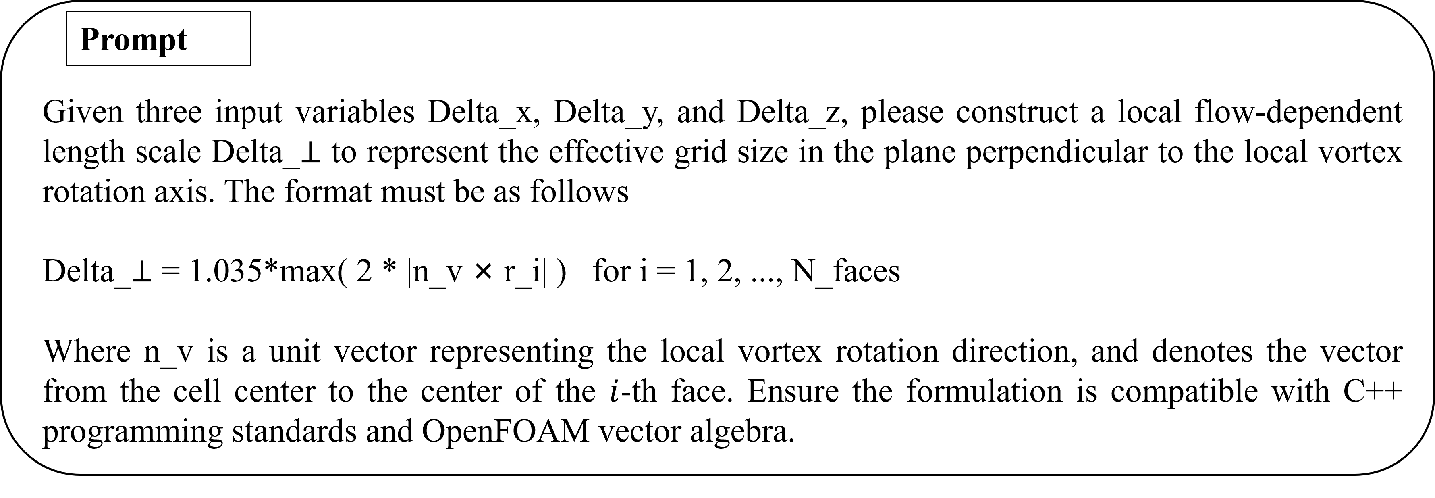}
  \caption{The user-defined prompt to initiate the DEScover workflow for length scale discovery.}
  \label{fig:prompt}
\end{figure}

\textbf{Discovery Agent}. To initiate the DEScover workflow, a user-defined prompt is provided to the Discovery Agent. In the prompt, the length scale is defined as the grid size in the direction orthogonal to the vortex axis, i.e., equation~\eqref{eq:delta_perp}. Figure~\ref{fig:prompt} shows the specific prompt content. Based on the prompt, the Discovery Agent performs literature analysis and physical reasoning (figure~\ref{fig:AutoTM}) to identify methods that determine the local vortex-axis direction $\mathbf{n}_v$ from the resolved velocity field, including vortex-identification criteria that were not originally formulated for directional identification. Each method is formulated as a candidate length scale of the form of equation~\eqref{eq:delta_perp}, together with the fallback treatment required in regions where the vortex axis is undefined. In subsequent iterations of the closed loop, the feedback from the Evaluation Agent (\S\,\ref{sec:closed_loop}) is appended to the prompt, so that the Discovery Agent can refine the existing candidates or propose new ones.

\textbf{Review Agent}. After the Discovery Agent proposes new length-scale candidates, the Review Agent evaluates them before implementation in OpenFOAM. The Review Agent examines and refines each candidate based on four criteria. (1) \textit{Internal logical consistency}. This criterion ensures that each candidate is logically consistent throughout the proposed formulation. For example, if a proposal uses the maximum eigenvalue, the corresponding formulation should also use the maximum eigenvalue rather than incorrectly referring to the minimum eigenvalue. Any inconsistency is identified and corrected before further implementation. (2) \textit{Dimensional homogeneity}. This criterion ensures that every proposed expression is dimensionally consistent. The dimensions on the left-hand side and right-hand side of each equation must be identical. Any expression that violates this requirement is considered invalid and must be revised to satisfy dimensional consistency. (3) \textit{Locality}. vortex axis direction is a pointwise function of the local velocity gradient tensor, $\mathbf{n}_v=f(\nabla\boldsymbol{u})$, independent of any spatial, temporal or global averages.The subgrid length scale represents the local resolution capability of the grid. A non-local definition would make it depend on the averaging procedure, on the choice of averaging domain and on the convergence of flow statistics, and $\mathbf{n}_v$ would be ill-defined during the initial transient of a simulation. Locality also lets each cell be evaluated independently, so the formulation requires no global reductions or additional stored mean fields and is directly compatible with parallel computation on arbitrary unstructured meshes. (4) \textit{Physical uniqueness}. vortex axis direction is the intrinsic rotation axis (or swirling-plane normal) defined by a single vortex identification method, rather than an ad hoc combination of directions from different methods. Ad hoc combinations, weightings or interpolations of directions obtained from different methods are excluded. The definition of $\Delta_{\perp}$ rests on the premise that $\mathbf{n}_v$ is a physical rotation axis, because $\Delta_{\perp}$ is meant to measure the grid spacing in the plane perpendicular to that axis. A blended direction generally corresponds to no physical axis. It also introduces non-unique modelling choices, such as the form of the weighting function or switching thresholds, that cannot be justified on physical grounds.

\subsubsection{Code Implementation in OpenFOAM}
\label{sec:openfoam}
Rather than requiring the LLM to generate OpenFOAM code from scratch, we supply an
existing IDDES length scale implementation as a reference template.  The new length scale
expression is implemented by inheriting from this base class, leveraging both the
object-oriented structure of C++ and the modular turbulence model framework of OpenFOAM.
This design significantly reduces the implementation burden and minimises the risk of code
hallucination, as the LLM need only modify the length scale computation within a
well-defined, pre-validated code structure. Specifically, for each candidate, the LLM is prompted to generate the C++ code that computes the vortex-axis direction $\mathbf{n}_v$ from the velocity gradient tensor $\nabla\mathbf{U}$, including the fallback treatment, and evaluates equation~\eqref{eq:delta_perp} using the cell centroid-to-face-centroid vectors $\mathbf{r}_f$. The generated code is inserted into the designated length-scale function of the template class. A Python validation script checks for syntactic correctness and the absence of division-by-zero or imaginary-valued sub-expressions, and a dry-run compilation is then performed with the standard OpenFOAM \texttt{wmake} build system before the code is used in the CFD simulations.

This approach is consistent with the natural-language-to-intermediate-representation
philosophy advocated by \citet{shenoy2026natural}, where the language model
infers high-level intent rather than generating unconstrained executable code.  By confining
the LLM to a well-defined function slot within a validated template, the risk of silent semantic errors and hallucinated solver directives is greatly reduced. However, code that compiles and runs may still implement a wrong formula. To detect such errors, an additional consistency check is performed after the code is implemented. The generated code is compared with the formula in the corresponding proposal file, term by term. The check covers the definition of the vortex-axis direction $\mathbf{n}_v$, the fallback treatment, and the evaluation of equation~\eqref{eq:delta_perp}. Any mismatch is returned to the Code Implementation Agent for correction. Only the code that passes this check is used in the CFD simulations.

\subsubsection{Hierarchical multi-case simulation module of DEScover}
\label{sec:hierarchical}
In this task, a hierarchical, multi-case CFD simulation module is established in DEScover. The module consists of several benchmark cases with increasing levels. For a specified turbulence modeling task, a case sequence $[c_1, c_2, c_3, ..., c_n]$ should be provided. The computational cost of the cases should generally increase along the sequence. The cases should be carefully selected to evaluate different aspects of candidate models. Typical cases include a flat plate for assessing attached-flow prediction, a backward-facing step for evaluating flow separation, a free shear layer for assessing the K--H instability, and an airfoil for evaluating curved-surface and high-Reynolds-number effects. These benchmark cases are organized into a hierarchical evaluation framework. A candidate model is first evaluated using the case $c_1$. Only candidates that satisfy the $c_1$ evaluation criteria are allowed to proceed to the next level, namely the case $c_2$. Models that fail at any level are excluded from subsequent evaluations. Therefore, a successful candidate model must sequentially pass the cases $[c_1, c_2, c_3, ..., c_n]$. This hierarchical mechanism provides an efficient assessment of the generalization capability of candidate models. In addition, it significantly reduces the computational cost of CFD simulations, since models that fail at an early stage are not evaluated in the subsequent, more computationally expensive cases.

\subsubsection{The closed-loop capability of DEScover}
\label{sec:closed_loop}
In this task, the closed-loop capability of the DEScover framework is developed. The closed-loop process is achieved by introducing an Evaluation Agent. If all candidates fail to satisfy the CFD evaluation module, the Evaluation Agent provides detailed feedback and instructs the Discovery Agent to refine the prompt. More specifically, the Evaluation Agent generates a markdown file named \texttt{comments.md}. The first row of \texttt{comments.md} records the evaluation result, indicating either \textbf{Pass} or \textbf{Fail}. The remaining rows provide detailed failure comments based on the CFD evaluation criteria. This feedback enables the Discovery Agent to generate more reliable and physically feasible candidate ideas in the next iteration. The updated prompt is then used to generate new candidate ideas. This iterative process continues until a candidate satisfies the CFD evaluation criteria. Once a candidate passes the CFD evaluation criteria, the Evaluation Agent terminates the iterative loop and outputs the best candidate. No further refinement is required.

\subsection{Numerical Methodology}
\label{sec:numerics}

\subsubsection{Solver and Time Integration}
\label{sec:solver}
All computations were carried out using the open-source finite-volume solver OpenFOAM v2312.
The incompressible Navier--Stokes equations are solved with the transient solver \texttt{pimpleFoam} together with the SA-based IDDES model (\texttt{SpalartAllmarasIDDES}). The IDDES filter width follows equation~\eqref{eq:iddes_delta} with $C_w=0.15$, and the outer length scale $\Delta_{\text{outer}}$ is selected at run time, which allows each candidate length scale to be tested without modifying the remaining parts of the model. Pressure--velocity coupling is handled by the PIMPLE algorithm with one non-orthogonal corrector per time step. The pressure equation is solved with the geometric--algebraic multigrid (GAMG) solver with a relative tolerance of 0.05 (0 at the final corrector), and the momentum and $\tilde{\nu}$ equations are solved with preconditioned bi-conjugate gradient methods (DILU preconditioner). The case-specific numbers of outer and pressure correctors, under-relaxation factors, and solver tolerances are listed in table~\ref{tab:time_steps}. Time integration is performed with the second-order three-time-level implicit backward differencing scheme:
\begin{equation}
  \frac{\partial \phi}{\partial t} \approx \frac{3\phi^{n+1} - 4\phi^n + \phi^{n-1}}{2\Delta t}
  \label{eq:bdf2}
\end{equation}
which is equivalent in formal order to the implicit second-order backward-Euler scheme
employed by \citet{shur2008hybrid} in their NTS finite-volume code.  A fixed time step is used in each case. The time steps range from $\Delta t = 5\times10^{-7}$~s for the free shear layer to $\Delta t = 1\times10^{-5}$~s for the flat plate, and are listed in table~\ref{tab:time_steps} together with the corresponding non-dimensional values $\Delta t\,U_{\text{ref}}/L_{\text{ref}}$.

\subsubsection{Spatial Discretisation}
\label{sec:spatial}
Gradients are computed with the second-order Gauss linear scheme, and the diffusive (viscous) terms are discretized with the Gauss linear scheme supplemented by explicit non-orthogonal correction. The wall distance required by the SA model is computed with the mesh-wave method. For the convective term of the momentum equation, the DEShybrid scheme of \citet{travin2002physical} is adopted. This scheme blends a
low-dissipation interpolant (Scheme~1) with a bounded, more dissipative interpolant (Scheme~2) through
a sensor function $\sigma$ that evaluates the local resolved-to-modelled turbulence ratio:
\begin{equation}
  \phi_f = (1 - \sigma)\,\phi_f^{(1)} + \sigma\,\phi_f^{(2)}, \qquad 0 \le \sigma \le 1
  \label{eq:deshybrid}
\end{equation}
where $\sigma \approx 0$ in vorticity-dominated LES regions and $\sigma \approx 1$ in
irrotational or wall-layer RANS regions.  The case-specific time step settings are listed in table~\ref{tab:time_steps}. The second-order central differencing scheme (\texttt{linear}) is used as Scheme~1, whereas the TVD-limited cubic scheme (\texttt{limitedCubic}, limiter coefficient 1) is used as Scheme~2. The sensor $\sigma$ is evaluated with $C_{\text{DES}} = 0.65$ and bounded between 0 and 1. The convective term of the $\tilde{\nu}$ equation is discretized with the bounded second-order \texttt{limitedLinear} scheme (limiter coefficient 1) in all cases.

\begin{table}
  \centering
  \caption{Case-specific time step and solver settings. $U_{\text{ref}}$ and $L_{\text{ref}}$ denote the reference velocity and length used to non-dimensionalize the time step.}
  \label{tab:time_steps}
  {
  \begin{tabular}{lcccc}
    \toprule
    Case & $\Delta t$ (s) & $U_{\text{ref}}$ (m/s) & $L_{\text{ref}}$ (m) & $\Delta t\,U_{\text{ref}}/L_{\text{ref}}$ \\[3pt]
    \midrule
    Flat plate           & $1\times10^{-5}$ & $U_\infty=68.67$          & $L=1$     & $6.9\times10^{-4}$ \\
    Backward-facing step & $5\times10^{-6}$ & $U_{\text{ref}}=11.3$     & $H=0.038$ & $1.5\times10^{-3}$ \\
    Free shear layer     & $5\times10^{-7}$ & $U_1=41.54$               & $L=1$     & $2.1\times10^{-5}$ \\
    Iced airfoil         & $2\times10^{-6}$ & $U_\infty=37.44$          & $c=1$     & $7.5\times10^{-5}$ \\
    \bottomrule
  \end{tabular}}
\par\end{table}

\section{Results and Discussion}
\label{sec:results}

This section presents the overall workflow, case-specific results, and physical interpretation of the different length scales. During the hierarchical multi-case simulations, some candidate models are discarded and therefore do not proceed to subsequent cases. In each case, only the candidates that enter this case are shown. The only exception is the baseline $\Delta_{\max}$. It is simulated manually in all cases after the workflow is completed, and it is shown as a reference.

\subsection{Overall Workflow}
\label{sec:workflow}

\subsubsection{Overall running process}
\label{sec:expressions}

\begin{figure}[hbt!]
  \centering
  \includegraphics[width=1.0\textwidth]{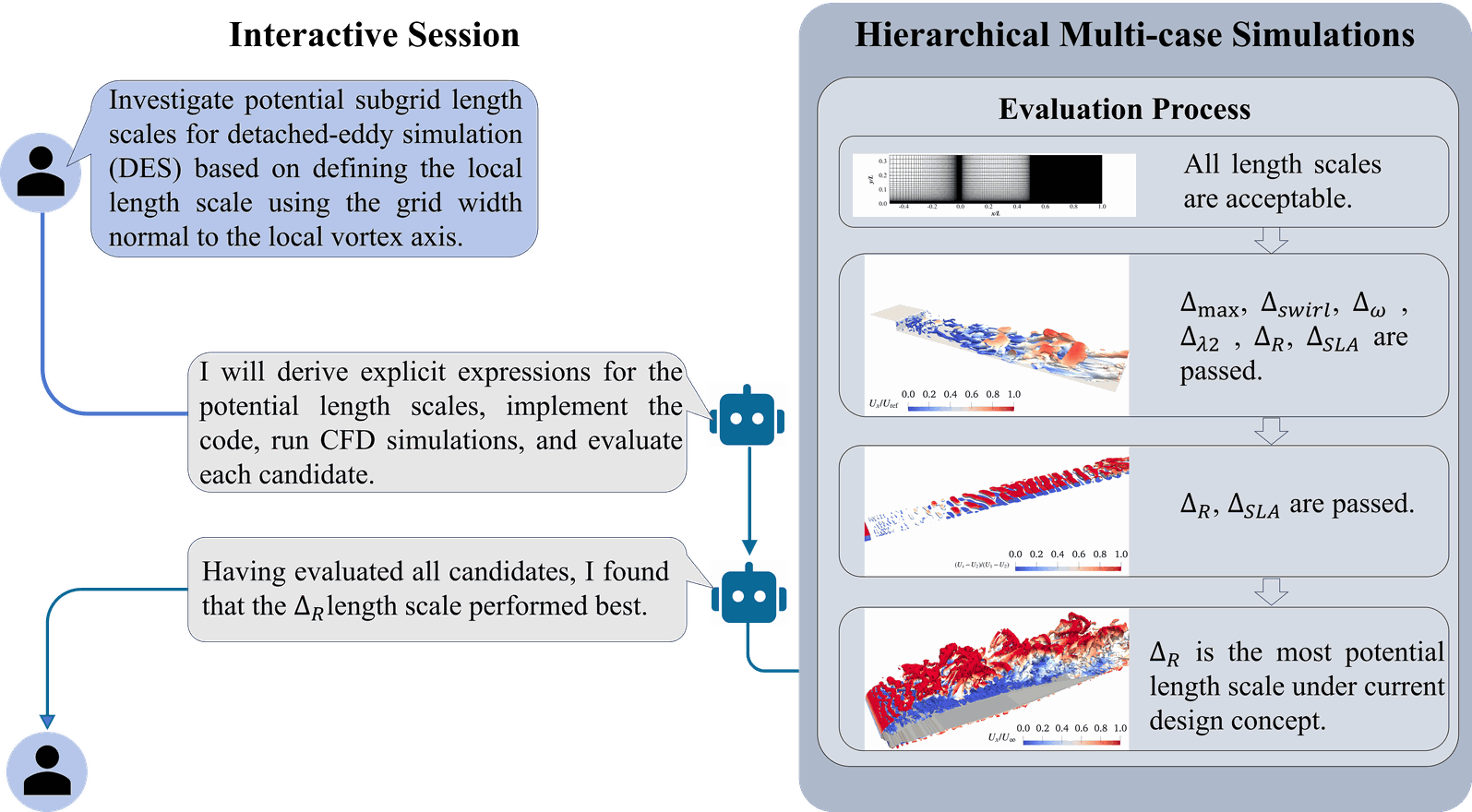}
  \caption{Interaction between the user and the DEScover framework in the subgrid length-scale discovery process.}
  \label{fig:interactive_framwork}
\end{figure}

Figure~\ref{fig:interactive_framwork} summarizes the interaction between the user and the DEScover framework in the present discovery task, together with the resulting evaluation process. The user provides only a high-level modeling concept in natural language, namely, defining the local length scale by the grid width normal to the local vortex axis. Based on this concept, the agents autonomously derive explicit expressions for the candidate length scales, implement them in OpenFOAM, perform the CFD simulations, and evaluate each candidate, without further human intervention. The right panel of figure~\ref{fig:interactive_framwork} shows how the candidates are progressively screened along the case sequence in table~\ref{tab:test_criteria}. All candidate length scales give acceptable predictions for the flat plate. The BFS and free-shear-layer cases, both dominated by the K--H instability of a shear layer, narrow the candidates down to $\Delta_R$ and $\Delta_{\text{SLA}}$. The iced-airfoil case finally identifies $\Delta_R$ as the most promising length scale under the current modeling concept, and the framework returns $\Delta_R$ to the user as the final output. The results supporting each screening step are presented in the following subsections.

All agents are driven by Claude Opus 5.5. The closed loop is executed for only one iteration, which demonstrates the effectiveness of the Discovery and Review Agents in identifying and screening candidate models. Human intervention is limited to writing the initial prompt and providing the case sequence and reference data. To assess the reproducibility of the LLM-based discovery, the Discovery, Review, and Code Implementation Agents are run five times with the same prompt. These three agents account for almost all of the token consumption of DEScover. The subsequent CFD simulations and evaluations are not repeated because of their computational cost. Table~\ref{tab:autotm_stats} lists the token consumption and wall-clock time of each agent, together with the success rate of each run. Among the five runs, four representative candidates, $\Delta_{\omega}$, $\Delta_{\text{swirl}}$, $\Delta_{\lambda_2}$, and $\Delta_{R}$, were proposed most frequently. These candidates have clear physical meanings and are derived from four distinct vortex-identification methods. The success rate (SR) is defined as 100\% when all four candidates are proposed by the Discovery Agent, pass the Review Agent, and are successfully implemented by the Code Implementation Agent.

\begin{table}
  \centering
  \caption{Reproducibility of DEScover: token consumption, wall-clock time and success rate over five runs. Tokens are given in millions (M) and include input, output and prompt-cache tokens; time is the wall-clock time.}
  \label{tab:autotm_stats}
  \setlength{\tabcolsep}{5pt}
  \begin{tabular}{lrrrrrrrrr}
    \toprule
    & \multicolumn{2}{c}{\begin{tabular}[b]{@{}c@{}}Discovery\\Agent\end{tabular}}
    & \multicolumn{2}{c}{\begin{tabular}[b]{@{}c@{}}Review\\Agent\end{tabular}}
    & \multicolumn{2}{c}{\begin{tabular}[b]{@{}c@{}}Code\\Implementation\\Agent\end{tabular}}
    & \multicolumn{3}{c}{Total} \\
    \cmidrule(lr){2-3} \cmidrule(lr){4-5} \cmidrule(lr){6-7} \cmidrule(lr){8-10}
    Run & Tokens & Time & Tokens & Time & Tokens & Time & Tokens & Time & SR \\
        & (M) & (min) & (M) & (min) & (M) & (min) & (M) & (min) & (\%) \\
    \midrule
    1 & 0.218 & 5.21 & 0.580 & 3.64 & 1.433 & 4.95 & 2.230 & 13.79 & 100 \\
    2 & 0.379 & 6.32 & 0.813 & 5.92 & 1.233 & 6.09 & 2.425 & 18.34 & 100 \\
    3 & 0.456 & 13.27 & 0.436 & 4.70 & 2.204 & 14.77 & 3.096 & 32.74 & 100 \\
    4 & 0.448 & 9.99 & 0.636 & 4.59 & 2.106 & 19.43 & 3.191 & 34.01 & 100 \\
    5 & 0.499 & 11.09 & 0.723 & 6.67 & 1.462 & 7.34 & 2.684 & 25.10 & 100 \\
    \midrule
    Mean & 0.400 & 9.18 & 0.638 & 5.11 & 1.687 & 10.51 & 2.725 & 24.79 & 100 \\
    Std.\ dev. & 0.111 & 3.35 & 0.143 & 1.20 & 0.437 & 6.29 & 0.416 & 8.82 & 0 \\
    Std.\ dev./mean (\%) & 27.6 & 36.5 & 22.5 & 23.4 & 25.9 & 59.8 & 15.3 & 35.6 & 0 \\
    \bottomrule
  \end{tabular}
\par\end{table}

\subsubsection{Representative LLM generated Candidates}
\label{sec:candidates}

Not all candidates generated by the Discovery Agent pass the review stage. Candidates that violate any of the four criteria are rejected by the Review Agent. For example, one candidate defines the vortex-axis direction from the relative vorticity,
$\mathbf{n}_v = (\boldsymbol{\omega} - \boldsymbol{\omega}_{\mathrm{ref}})/|\boldsymbol{\omega} - \boldsymbol{\omega}_{\mathrm{ref}}|$,
following the instantaneous vorticity deviation of \citet{haller2016defining}. It is rejected because it violates criterion (3), locality. The reference vorticity $\boldsymbol{\omega}_{\mathrm{ref}}$ is obtained from a spatial, planar, or temporal average. Thus, $\mathbf{n}_v$ is not a pointwise function of the local velocity-gradient tensor and depends on the averaging procedure and the convergence of the flow statistics. Another candidate defines $\mathbf{n}_v$ as a rotation-weighted combination of the Liutex and vorticity directions,
$\mathbf{n}_v \propto w\mathbf{r} + (1-w)\boldsymbol{\omega}/|\boldsymbol{\omega}|$,
with $w=R/|\boldsymbol{\omega}|$. It is rejected because it violates criterion (4), physical uniqueness. This interpolation combines directions from two different vortex-identification methods without a unique physical basis. The resulting direction does not represent a specific physical rotation axis. Consequently, the swirling-plane grid spacing $\Delta_{\perp}$ constructed from this direction loses its physical meaning. Table \ref{tab:vortex_proposals} provides a systematic comparison of four distinct proposals for identifying the local vortex rotation axis $\mathbf{n}_v$. The complete formulations of Proposals 1--4, including the fallback strategies and the
intermediate tensor derivations, are provided in Appendix~\ref{app:proposals} for
reference. The OpenFOAM implementations of these length scales are publicly available at \url{https://github.com/iDesign-Lab/DEScover}.

\begin{table}
  \centering
  \caption{Comparison of different rotation axis identification candidates.}
  \label{tab:vortex_proposals}
  \begin{tabularx}{\textwidth}{@{}>{\raggedright\arraybackslash}p{5.6em}*{4}{>{\raggedright\arraybackslash}X}@{}}
    \toprule
    Feature & Proposal 1:\newline Vorticity & Proposal 2:\newline $\lambda_{ci}$ eigenvector & Proposal 3:\newline Rortex/Liutex & Proposal 4:\newline $S^2 + \Omega^2$ axis \\[3pt]
    \midrule
    Length scale & $\Delta_{\omega}$ & $\Delta_{\text{swirl}}$ & $\Delta_{R}$ & $\Delta_{\lambda_2}$ \\
    \addlinespace
    Definition & $\mathbf{n}_v = \boldsymbol{\omega} / \|\boldsymbol{\omega}\|$ & $\mathbf{n}_v = \mathbf{e}_r / \|\mathbf{e}_r\|$ & $\mathbf{n}_v = \text{sign}(\boldsymbol{\omega} \cdot \mathbf{e}_r) \mathbf{e}_r$ & $\mathbf{n}_v = \mathbf{v}_1$ \\
    \addlinespace
    Physical Basis & Antisymmetric part of $\nabla \mathbf{u}$ & Eigen-structure of velocity gradient & Vorticity decomposition (Rot. + Shear) &  eigenvector associated with the largest eigenvalue, $\lambda_2$ criterion (Pressure Hessian) \\
    \addlinespace
    Shear Immunity & Poor (Contaminated) & Good (Swirl only) & Excellent (Pure rotation) & Good (Physical) \\
    \addlinespace
    Stability & Extremely High & Medium & Medium & Extremely High \\
    \addlinespace
    Computational Cost & Extremely Low & Medium & High & Medium \\
    \addlinespace
    Fallback & $\|\boldsymbol{\omega}\| \approx 0$ & Required if $\lambda_{ci} = 0$ & Required if $R \le 0$ & Not Required \\
    \bottomrule
  \end{tabularx}
\par\end{table}

\subsubsection{Test cases and evaluations}
\label{sec:evaluation}

The candidate length scales are evaluated in the hierarchical multi-case CFD simulation module described in \S\,\ref{sec:hierarchical}. The case sequence consists of four cases, arranged in order of increasing flow complexity and computational cost. (1) The flat-plate boundary layer of \citet{gritskevich2012development}. This case evaluates whether a candidate length scale provides boundary-layer protection similar to that of $\Delta_{\max}$ in attached flow. (2) The backward-facing step (BFS) of \citet{vogel1985combined} at $Re_H = 28{,}000$, based on the step height $H$. This case assesses whether a candidate length scale can alleviate the delayed onset of the Kelvin--Helmholtz (K--H) instability in the shear layer separating from the step edge. (3) The turbulent free shear layer of \citet{delville1995decomposition}, with freestream velocities of $U_1=41.54$~m/s and $U_2=22.40$~m/s in the upper and lower streams, respectively. This flow is dominated by the K--H instability without wall-bounded separation. It evaluates whether a candidate length scale can reduce the delayed onset of the K--H instability and the subsequent shear-layer roll-up, which are commonly observed with the existing length scale $\Delta_{\max}$. (4) The GLC305 airfoil with the 944 glaze-ice shape at $Re_c=3.5\times10^6$, $M=0.12$, and $\alpha=6^\circ$, measured by \citet{broeren2006flowfield}. This case represents a high-Reynolds-number engineering flow and provides the most challenging test of the generalization capability of the candidate length scales.

Table~\ref{tab:test_criteria} summarizes the flow conditions, reference data, and evaluation criteria of each case. At each level, the simulation results are compared with the reference data to quantify the errors, and the candidate length scales are ranked accordingly. Because all candidates give acceptable predictions for the flat plate, they are all retained at the first level. At each subsequent level, the two lowest-ranked length scales are discarded and excluded from the next test case, until the best-performing length scale is identified in the final case. The candidates retained after each case are listed in the last column of table~\ref{tab:test_criteria}. In this way, the number of candidates decreases as the computational cost of the cases increases. The computational setup and detailed results of each case are presented in \S\,\ref{sec:test_cases}.

\begin{table}
  \centering
  \caption{Flow conditions, reference data, evaluation criteria, and retained candidates for the test cases in the hierarchical multi-case simulation module.}
  \label{tab:test_criteria}
  \begin{tabularx}{\textwidth}{@{}>{\raggedright\arraybackslash}p{5.2em}*{4}{>{\raggedright\arraybackslash}X}@{}}
    \toprule
    Cases & Flow conditions & Reference data sources &  Evaluation criteria & Passed candidates \\[3pt]
    \midrule

    Flat plate & $Re_L=1.0 \times 10^7$ based on the length of the plate $L$ & \citet{gritskevich2012development} & skin friction $C_f$, eddy viscosity $\mu_t / \mu$ & $\Delta_{\text{max}}$, $\Delta_{\omega}$, $\Delta_{\text{swirl}}$, $\Delta_{\lambda_2}$, $\Delta_{R}$, $\Delta_{\text{SLA}}$ \\
    \addlinespace

    Backward-facing step & $Re_H = 28{,}000$, based on step height $H$ & \citet{vogel1985combined}  & skin friction $C_f$,  eddy viscosity, Mean streamwise velocity profiles, streamwise rms velocity fluctuation $u_{\mathrm{rms}}$  & $\Delta_{\omega}$, $\Delta_{\lambda_2}$, $\Delta_{R}$, $\Delta_{\text{SLA}}$ \\
    \addlinespace

    Free shear layer & $U_1=41.54$~m/s and $U_2=22.40$~m/s in the upper and lower streams & \citet{delville1995decomposition}  & Mean streamwise velocity profiles  & $\Delta_{R}$, $\Delta_{\text{SLA}}$ \\
    \addlinespace

    Iced airfoil & $Re_c=3.5 \times 10^6$, based on chord $c$ & \citet{broeren2006flowfield}  & Mean pressure coefficient $C_p$ & $\Delta_{R}$ \\
    \bottomrule
  \end{tabularx}
\par\end{table}

The ranking at each level is based on the quantitative error metrics summarized in table~\ref{tab:error_metrics}. For a quantity $\phi$ compared with the reference data $\phi_{\text{ref}}$ at $N$ measurement points, the relative $L_2$ error is defined as
\begin{equation}
\varepsilon_{\phi} = \sqrt{\frac{\sum_{i=1}^{N} \left(\phi_i - \phi_{\text{ref},i}\right)^2}{\sum_{i=1}^{N} \phi_{\text{ref},i}^2}}.
\label{eq:l2_error}
\end{equation}
For cases involving multiple evaluation metrics, equal weights are assigned to all metrics. For example, the backward-facing step case is evaluated based on the errors in skin friction coefficient $C_f$, mean velocity profiles, and streamwise rms velocity fluctuations $u_{\mathrm{rms}}$, with equal weights assigned to these three metrics. The reattachment length $x_r/H$ is also compared with the experimental measurement.

\begin{table}
  \centering
  \caption{Quantitative errors of the candidate length scales at each level of the hierarchical multi-case simulation module. ``--'' denotes a candidate that was discarded at a previous level.}
  \label{tab:error_metrics}
  \small
  \begin{tabular}{lcccccc}
    \toprule
    Length scale & Flat plate & \multicolumn{3}{c}{Backward-facing step} & Free shear layer & Iced airfoil \\
    \cmidrule(lr){2-2}\cmidrule(lr){3-5}\cmidrule(lr){6-6}\cmidrule(lr){7-7}
     & $\varepsilon_{C_f}$ & $\varepsilon_{C_f}$ & $\varepsilon_{U}$ & $\varepsilon_{u_{\mathrm{rms}}}$ & $\varepsilon_{U}$ & $\varepsilon_{C_p}$ \\
    \midrule
    $\Delta_{\max}$       & 0.0617 & 1.046 & 0.285 & 0.555 & -- & -- \\
    $\Delta_{\text{swirl}}$ & 0.0614 & 1.023 & 0.269 & 0.515 & -- & -- \\
    $\Delta_{\omega}$     & 0.0611 & 0.798 & 0.201 & 0.418 & 0.259 & -- \\
    $\Delta_{\lambda_2}$  & 0.0612 & 0.864 & 0.246 & 0.404 & 0.262 & -- \\
    $\Delta_{\text{SLA}}$ & 0.0633 & 0.357 & 0.111 & 0.194 & 0.163 & 0.365 \\
    $\Delta_{R}$          & 0.0639 & 0.294 & 0.101 & 0.197 & 0.161 & 0.359 \\
    \bottomrule
  \end{tabular}
\par\end{table}

\subsection{Case-specific results}
\label{sec:test_cases}

\subsubsection{Flat plate}
\label{sec:Flat plate}

\begin{figure}[hbt!]
  \centering
  \begin{subfigure}[b]{0.8\textwidth}
    \centering
    \includegraphics[width=\textwidth]{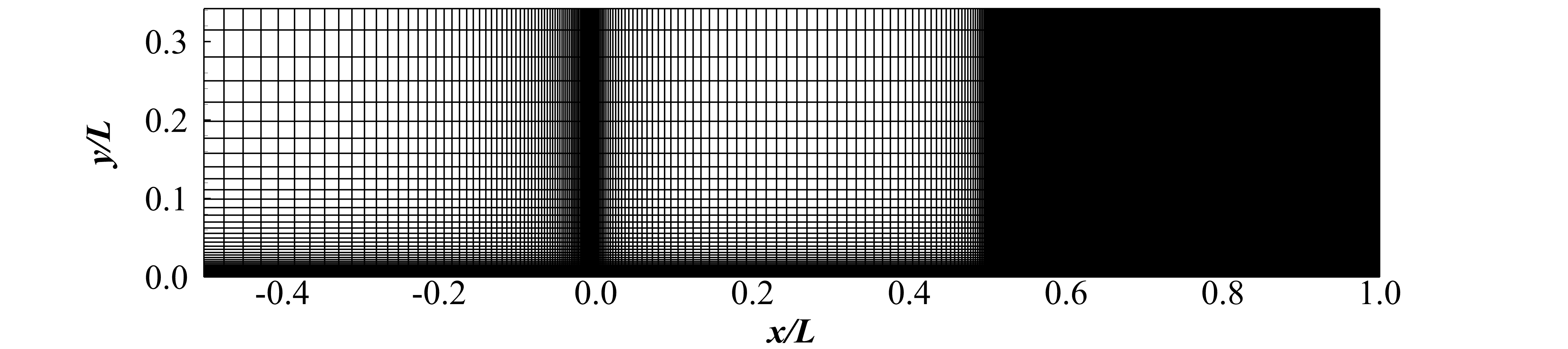}
    \caption{Full streamwise computational domain and grids.}
    \label{fig:plate_grid_a}
  \end{subfigure}
  \\ %
  \vspace{0.5cm} %

  \begin{subfigure}[b]{0.8\textwidth}
    \centering
    \includegraphics[width=\textwidth]{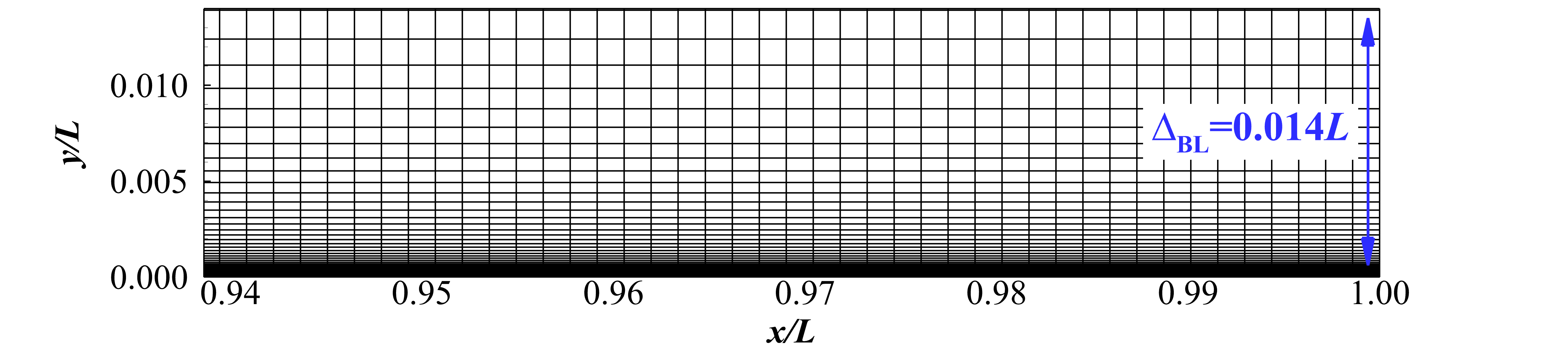} 
    \caption{Close-up view of the grids at outlet.}
    \label{fig:plate_grid_b}
  \end{subfigure}

  \caption{Computational domain and grids for flat plate: (a) full streamwise grid; (b) close-up view at outlet.}
  \label{fig:plate_grid}
\end{figure}

The first test case is a flat plate. It is used to assess whether the candidate length scales can predict acceptable boundary layer flow. The Reynolds number $Re_L$ based on the freestream velocity and the length of the plate $L$ is $1.0 \times 10^7$, and the Mach number is $0.2$. This case was also investigated by \citet{gritskevich2012development} using the original IDDES method. The methods applied and compared in this paper are SA turbulence model, the original IDDES and the IDDES with proposed length scales. The computational domain extends from $-0.5L$ to $1.0L$ in the streamwise direction. At the lower boundary, a symmetric boundary condition is implemented from $-0.5L$ to $0.0L$ upstream of the plate, and the plate is set as a no-slip wall. The upper boundary is located $1.0L$ away from the plate and is set as the far field boundary condition. Periodic conditions are implemented in the spanwise direction. The computational grid consists of $N_x=510$ cells in the streamwise direction, $N_y=100$ cells in the wall-normal direction, and $N_z=8$ cells in the spanwise direction.The first wall-normal cell distance is set as $5 \times 10^{-7}L$ to ensure $\Delta y^+ < 1$. The streamwise grid experiences a
sudden refinement at $x/L=0.5$, maintaining a grid spacing of $\Delta_x=\Delta_z=0.0014L$, which is approximately 0.1 times the boundary layer
thickness, as depicted in figure~\ref{fig:plate_grid}.

\begin{figure}[hbt!]
  \centering
  \includegraphics[width=0.6\textwidth]{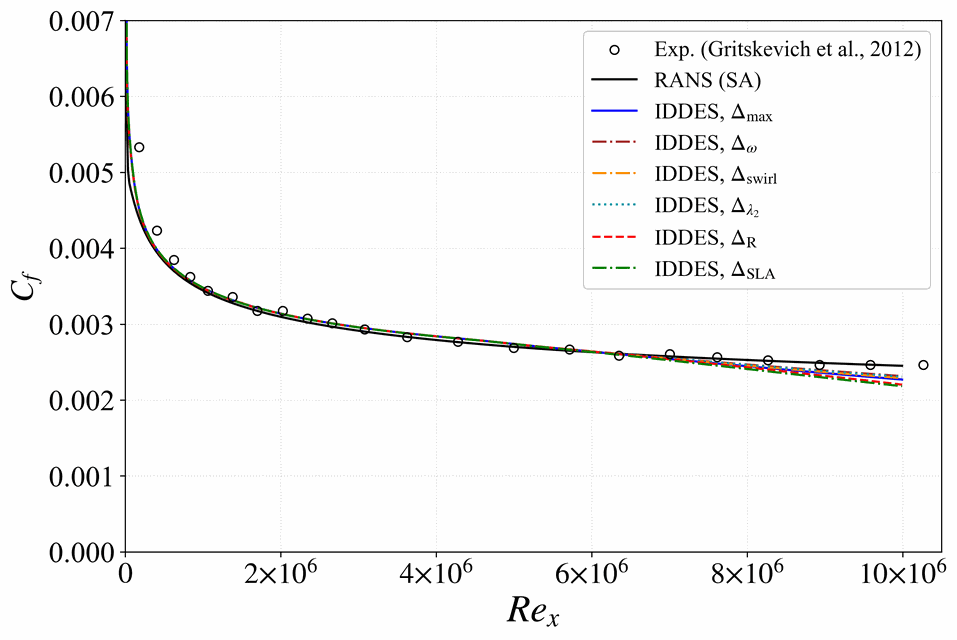}
  \caption{Comparison of the skin friction coefficient $C_f$ distributions obtained with different length scales (experimental data of \citealt{gritskevich2012development}) in flat plate.}
  \label{fig:plate_cf}
\end{figure}

\begin{figure}[hbt!]
  \centering
  \begin{subfigure}[b]{0.33\textwidth}
    \includegraphics[width=\textwidth]{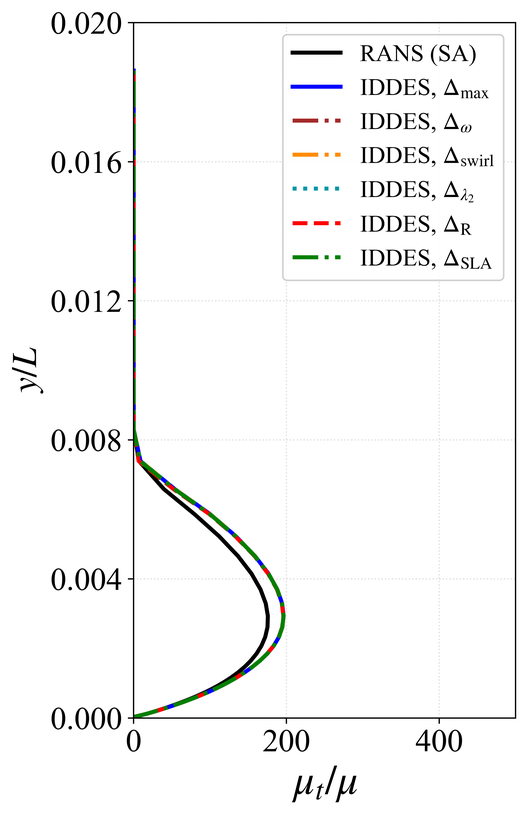}
    \caption{$Re_x=4.0 \times 10^6$}
  \end{subfigure}\hfill
  \begin{subfigure}[b]{0.33\textwidth}
    \includegraphics[width=\textwidth]{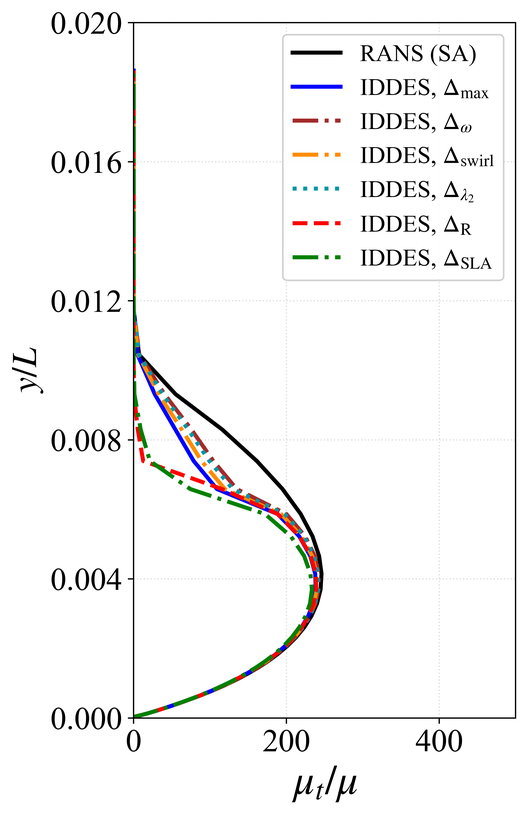}
    \caption{$Re_x=6.0 \times 10^6$}
  \end{subfigure}\hfill
  \begin{subfigure}[b]{0.33\textwidth}
    \includegraphics[width=\textwidth]{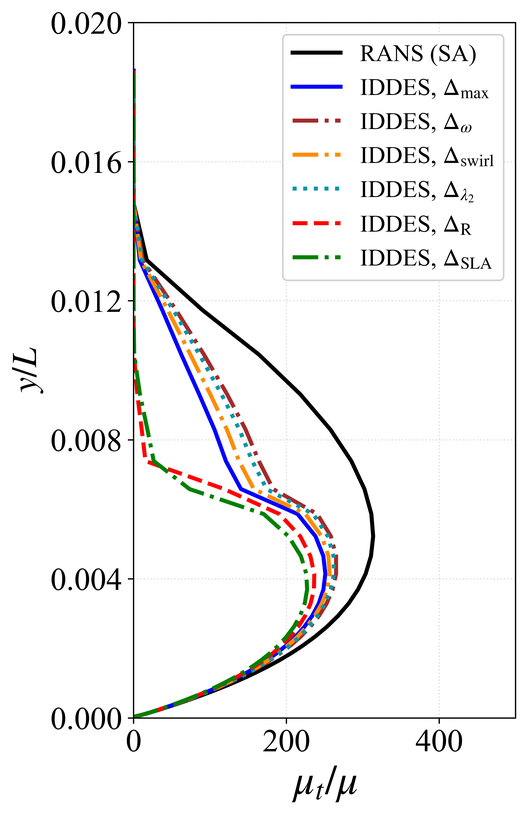}
    \caption{$Re_x=8.0 \times 10^6$}
  \end{subfigure}
  \caption{Dimensionless eddy viscosity $\mu_t/\mu$ profile at three locations: $Re_x=4.0 \times 10^6$, $Re_x=6.0 \times 10^6$ and $Re_x=8.0 \times 10^6$.}
  \label{fig:plate_nut}
\end{figure}

Figure~\ref{fig:plate_cf} presents the comparison of the skin friction coefficient $C_f$ distributions obtained with different length scales. RANS simulation can accurately reproduce the experimental results. The results obtained with $\Delta_{\text{max}}$, $\Delta_{\omega}$, $\Delta_{\text{swirl}}$, and $\Delta_{\lambda_2}$ are nearly identical, with only minor discrepancies near the trailing edge. $\Delta_{R}$ shows a slightly larger deviation from the reference result but remains comparable to $\Delta_{\text{SLA}}$. Overall, all candidate length scales provide satisfactory accuracy for the flat-plate case. Quantitatively, the relative error $\varepsilon_{C_f}$ of $\Delta_{R}$ is $6.39\%$, compared with $6.17\%$ for $\Delta_{\max}$ (table~\ref{tab:error_metrics}). The difference of about 0.2 percentage points is small compared with the improvements obtained in the separated-flow cases, and is confined to the downstream part of the plate ($Re_x \gtrsim 7\times10^6$), where all IDDES solutions slightly under-predict $C_f$. The slightly larger deviation of $\Delta_R$ is attributed to the fallback treatment of Proposal~3 (Appendix~\ref{app:proposals}). In an attached boundary layer, the velocity gradient tensor is dominated by the mean shear $\partial U/\partial y$, for which all eigenvalues of $\nabla\mathbf{u}$ are real. No vortex axis is then identified, and $\Delta_R$ vanishes. In the outer part of the boundary layer, where the branch $f_d<\varepsilon$ of equation~\eqref{eq:cdes_amd_lim} is active, $\Delta_{\text{outer},lim}$ therefore reduces to $C_{lim}V^{1/3}$, which is smaller than $\Delta_{\max}$. This reduces the eddy viscosity in the outer part of the boundary layer, as shown by the earlier decay of $\mu_t/\mu$ with $\Delta_R$ for $y/L \gtrsim 0.007$ in figure~\ref{fig:plate_nut}. The inner part of the boundary layer remains shielded, so the effect on $C_f$ is small.

Figure~\ref{fig:plate_nut} shows the dimensionless eddy viscosity $\mu_t/\mu$ profile at three different streamwise locations: $Re_x=4.0 \times 10^6$, $Re_x=6.0 \times 10^6$ and $Re_x=8.0 \times 10^6$. The wall-normal distance is normalized by the plate length $L$ instead of the boundary layer thickness because the boundary layer thickness varies along the streamwise direction. The results obtained with $\Delta_{\text{max}}$, $\Delta_{\omega}$, $\Delta_{\text{swirl}}$, and $\Delta_{\lambda_2}$ are similar. $\Delta_{R}$ shows a larger deviation from the reference result but remains comparable to $\Delta_{\text{SLA}}$.

\subsubsection{Backward-Facing Step}
\label{sec:Backward-Facing Step}

\begin{figure}[hbt!]
  \centering
  \includegraphics[width=0.9\textwidth]{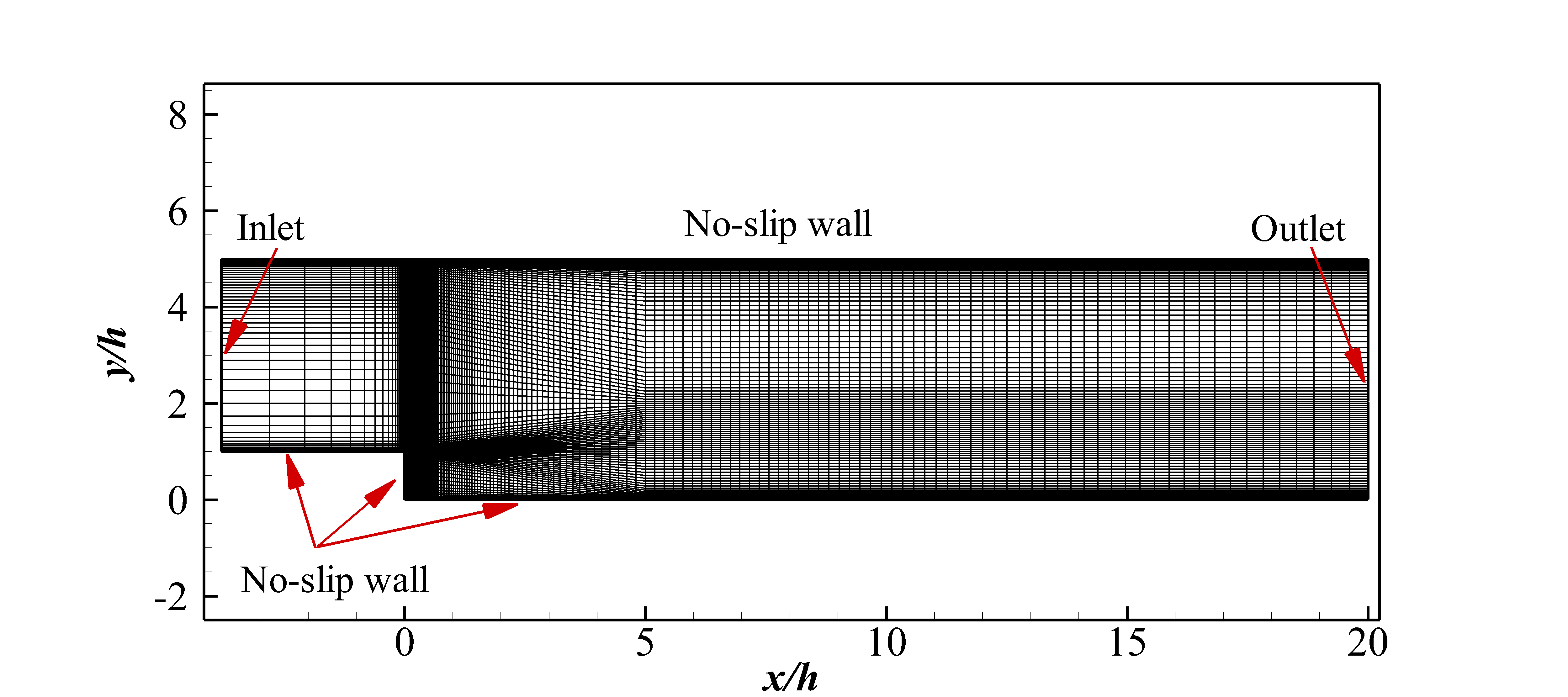}
  \caption{Computational domain and grids for backward-facing step.}
  \label{fig:bfs_grid}
\end{figure}

The second case in the DEScover case sequence is the backward-facing step (BFS) at $Re_H = 28{,}000$, based on the step height $H$, measured by \citet{vogel1985combined}. This case is selected because of its well-documented grey-area behavior in the separating shear layer and the availability of detailed experimental data for $C_f$, mean velocity, and Reynolds stresses.  Figure~\ref{fig:bfs_grid} shows the computational domain for backward facing step. The channel expansion
ratio is 5/4, and the incoming boundary-layer thickness in the experiment \citep{vogel1985combined} is turbulent with
a thickness of $1.07H$. The computational domain extended from $x/H = -3.8$ to $x/H = 20$ in the streamwise direction, with $x = 0$ corresponding to the step location, and $4H$ in the spanwise
direction.  Periodic boundary conditions were applied in the spanwise direction, a constant
pressure condition was specified at the outlet, and steady-state RANS profiles were imposed
at the inlet---with unsteadiness arising from the inherent flow instability past the step. The computational grid comprised 2.56~million hexahedral cells (2.62~million nodes),
providing near-wall resolution in wall units of $y^+ < 1$.  A fixed time step of $\Delta t = 5\times10^{-6}$~s was used (\S\,\ref{sec:numerics}). The maximum Courant number of each BFS simulation is about 1.1. The number of cells in the spanwise direction was 79.

\begin{figure}[hbt!]
  \centering
  
  \begin{subfigure}[b]{0.45\textwidth}
    \includegraphics[width=\textwidth]{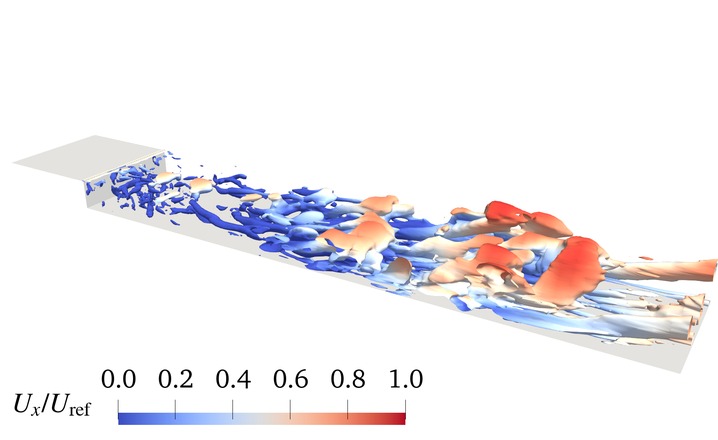}
    \caption{$\Delta_{\max}$}
  \end{subfigure}
  \hfill
  \begin{subfigure}[b]{0.45\textwidth}
    \includegraphics[width=\textwidth]{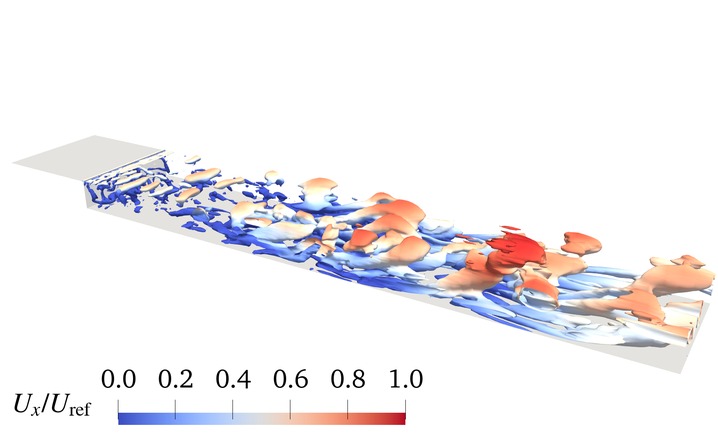}
    \caption{$\Delta_{\mathrm{\omega}}$}
  \end{subfigure}

  \vspace{0.5cm} %
  
  \begin{subfigure}[c]{0.45\textwidth}
    \includegraphics[width=\textwidth]{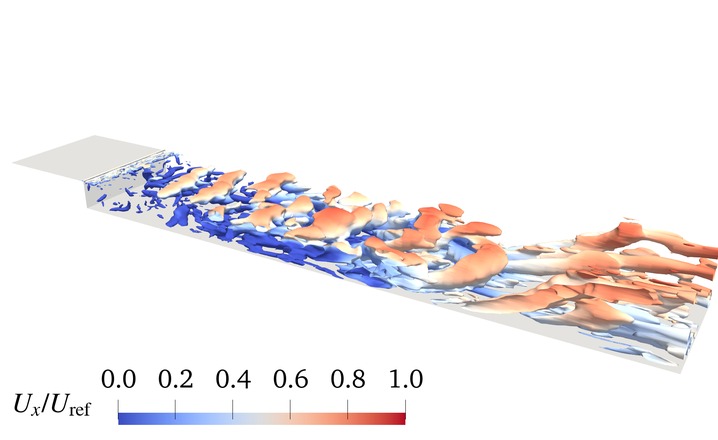}
    \caption{$\Delta_{\lambda_2}$}
  \end{subfigure}
  \hfill
  \begin{subfigure}[d]{0.45\textwidth}
    \includegraphics[width=\textwidth]{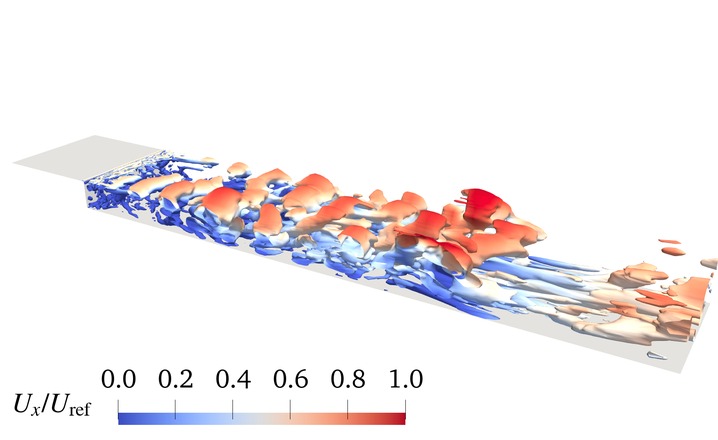}
    \caption{$\Delta_{\mathrm{swirl}}$}
  \end{subfigure}

  \vspace{0.5cm} %
  
  \begin{subfigure}[d]{0.45\textwidth}
    \includegraphics[width=\textwidth]{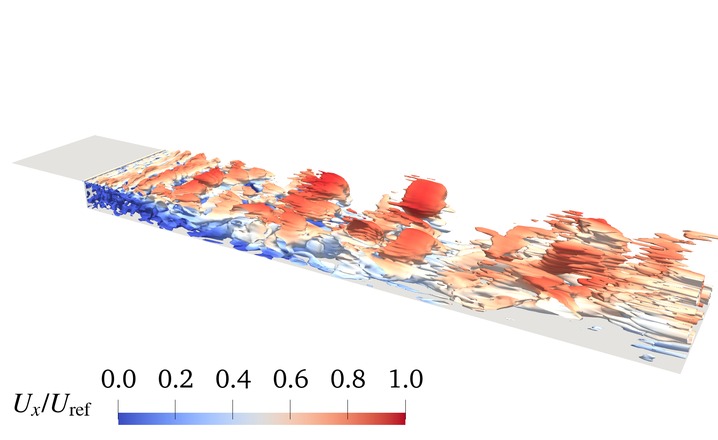}
    \caption{$\Delta_{\mathrm{R}}$}
  \end{subfigure}
  \hfill
  \begin{subfigure}[d]{0.45\textwidth}
    \includegraphics[width=\textwidth]{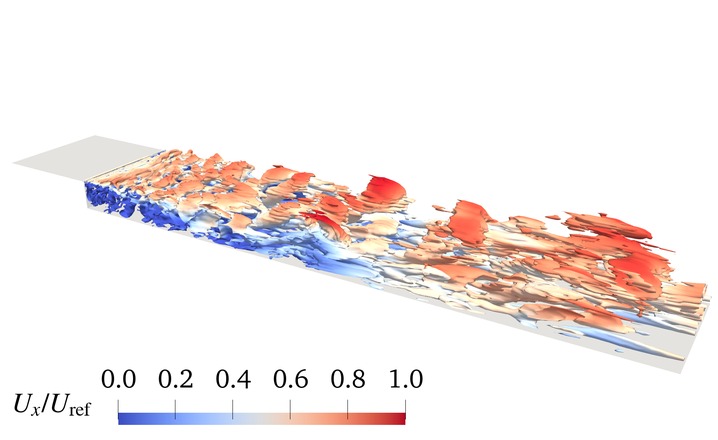}
    \caption{$\Delta_{\mathrm{SLA}}$}
  \end{subfigure}
  \caption{Different vortex evolution processes using different length scales in backward facing step. Isosurfaces of the dimensionless Q criterion, $Q(H/U_{\text{ref}})^2=0.05$, colored by $U_x/U_{\text{ref}}$.}
  \label{fig:bfs_Q}
\end{figure}

Figure~\ref{fig:bfs_Q} compares the vortex evolution predicted using different length scales in the backward-facing-step flow. The vortical structures are visualized by isosurfaces of the dimensionless Q criterion, $Q(H/U_{\text{ref}})^2=0.05$, colored by the dimensionless streamwise velocity ($U_x/U_{\text{ref}}$). The same isosurface value is used in all six panels. The $\Delta_{\text{max}}$ model predicts the largest separation region. $\Delta_{\omega}$, $\Delta_{\text{swirl}}$, and $\Delta_{\lambda_2}$ exhibit similar behavior to $\Delta_{\text{max}}$. In contrast, $\Delta_{\mathrm{R}}$ effectively reduces the delay of Kelvin--Helmholtz (K--H) instability and produces the smallest separation region. Its behavior is similar to that of $\Delta_{\text{SLA}}$.

\begin{figure}[hbt!]
  \centering
  \begin{subfigure}[b]{0.48\textwidth}
    \centering
    \includegraphics[width=\textwidth]{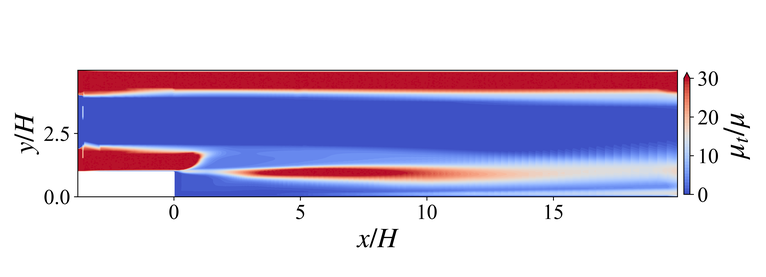}
    \caption{$\Delta_{\max}$}
    \label{fig:bfs_nut_a}
  \end{subfigure}%
  \hfill
  \begin{subfigure}[b]{0.48\textwidth}
    \centering
    \includegraphics[width=\textwidth]{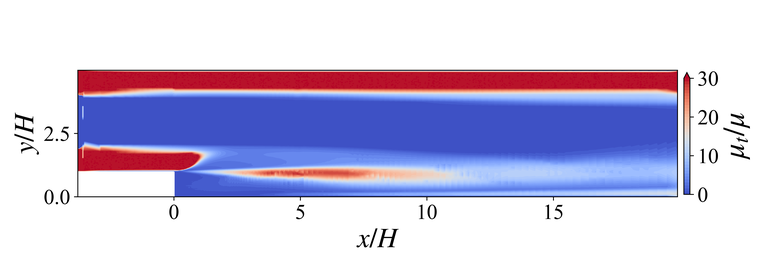}
    \caption{$\Delta_{\mathrm{\omega}}$}
    \label{fig:bfs_nut_b}
  \end{subfigure}

  \vspace{0.5cm} %

  \begin{subfigure}[b]{0.48\textwidth}
    \centering
    \includegraphics[width=\textwidth]{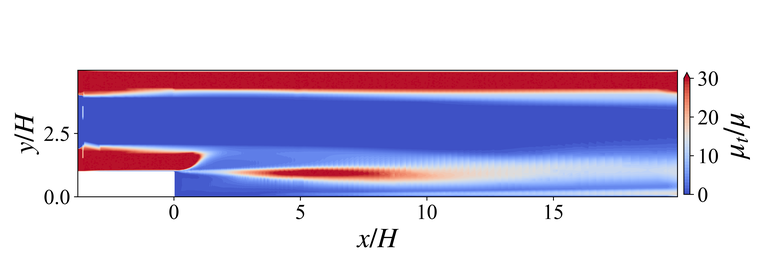}
    \caption{$\Delta_{\lambda_2}$}
    \label{fig:bfs_nut_c}
  \end{subfigure}%
  \hfill
  \begin{subfigure}[b]{0.48\textwidth}
    \centering
    \includegraphics[width=\textwidth]{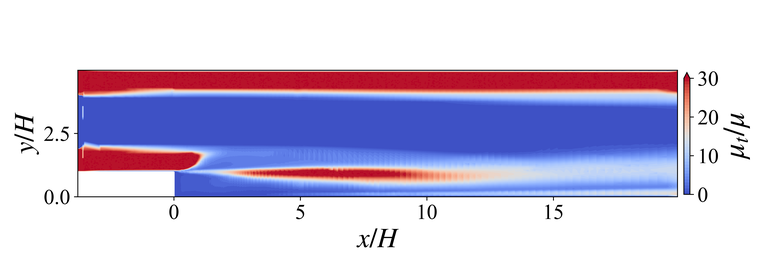}
    \caption{$\Delta_{\mathrm{swirl}}$}
    \label{fig:bfs_nut_d}
  \end{subfigure}

  \vspace{0.5cm}

  \begin{subfigure}[b]{0.48\textwidth}
    \centering
    \includegraphics[width=\textwidth]{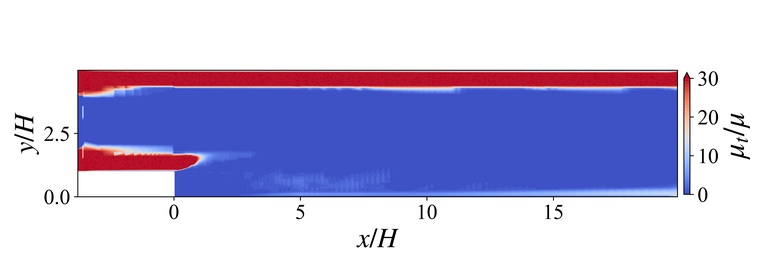}
    \caption{$\Delta_{\mathrm{R}}$}
    \label{fig:bfs_nut_e}
  \end{subfigure}%
  \hfill
  \begin{subfigure}[b]{0.48\textwidth}
    \centering
    \includegraphics[width=\textwidth]{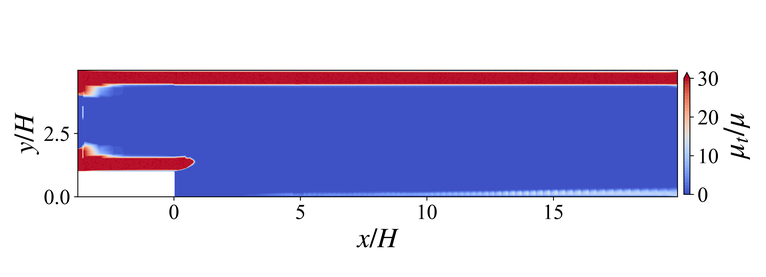}
    \caption{$\Delta_{\mathrm{SLA}}$}
    \label{fig:bfs_nut_f}
  \end{subfigure}

  \caption{Dimensionless eddy-viscosity ${\mu_{t}/\mu}$ distributions obtained using different length scales in backward facing step.}
  \label{fig:bfs_eddy_viscosity}
\end{figure}

Figure~\ref{fig:bfs_eddy_viscosity} shows the dimensionless mean eddy-viscosity ${\mu_{t}/\mu}$ distributions obtained using different length scales. Statistical averaging starts after the flow  passes through 7 computational domains. The subsequent 20 flow-through times are used to obtain the time-averaged flow field.. $\Delta_{\text{max}}$, $\Delta_{\omega}$, $\Delta_{\text{swirl}}$, and $\Delta_{\lambda_2}$ exhibit similar distributions. In contrast, $\Delta_{\mathrm{R}}$ produces a markedly different distribution, with very low eddy viscosity in the separated region. This behavior is similar to that obtained with $\Delta_{\text{SLA}}$.

\begin{figure}[hbt!]
  \centering
  \includegraphics[width=0.9\textwidth]{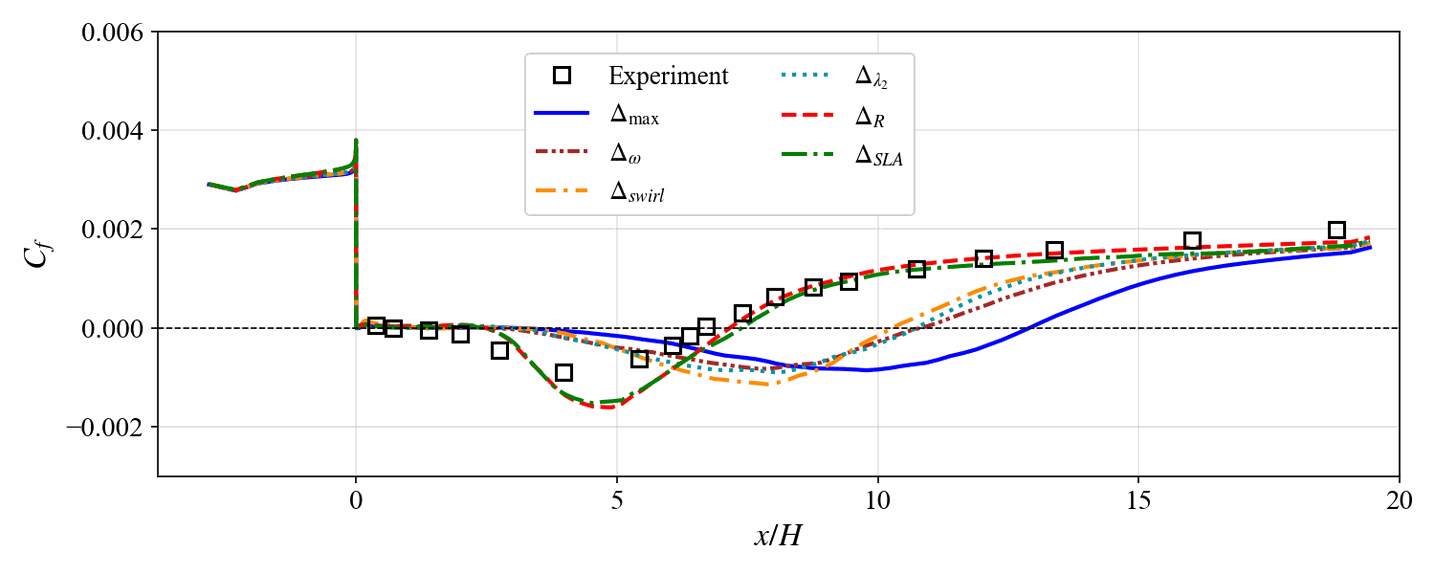}
  \caption{Distribution of the surface friction coefficient $C_f$ along the lower no-slip wall of the backward-facing step.}
  \label{fig:bfs_Cf}
\end{figure}

Figure~\ref{fig:bfs_Cf} presents the distribution of the surface friction coefficient $C_f$ along the lower no-slip wall of the backward-facing step. Among all considered length scales, $\Delta_{\text{max}}$ exhibits the largest discrepancy compared with the experimental data.  In contrast, the other length scales show improved agreement with the experimental results. The accuracy improves in the order of $\Delta_{\text{max}}$, $\Delta_{\text{swirl}}$, $\Delta_{\lambda_2}$, $\Delta_{\omega}$, $\Delta_{\text{SLA}}$, and $\Delta_{\mathrm{R}}$. Among these length scales, $\Delta_{R}$ and $\Delta_{\text{SLA}}$ show a significant improvement in prediction accuracy. This improvement is also reflected in the reattachment length $x_r/H$, defined as the location where $C_f$ changes sign from negative to positive downstream of the step (table~\ref{tab:bfs_reattachment}). The experimental reattachment length is $x_r/H=6.67$. $\Delta_{\max}$ predicts $x_r/H=12.50$, corresponding to a relative error of $87.4\%$, i.e., the recirculation region is nearly twice as long as in the experiment. $\Delta_{\text{swirl}}$, $\Delta_{\lambda_2}$, and $\Delta_{\omega}$ reduce the reattachment length to $11.40$, $10.79$, and $10.07$, respectively, but still over-predict it by more than $50\%$. In contrast, $\Delta_{\text{SLA}}$ and $\Delta_{R}$ predict $x_r/H=7.54$ and $7.13$, with relative errors of only $13.0\%$ and $6.9\%$. The ranking based on $x_r/H$ is identical to that based on $\varepsilon_{C_f}$ in table~\ref{tab:error_metrics}.

\begin{table}
  \centering
  \caption{Reattachment length $x_r/H$ of the backward-facing step predicted by the candidate length scales. $\delta_{x_r} = |x_{r} - x_{r,\text{exp}}|/x_{r,\text{exp}}$ denotes the relative error with respect to the experimental value.}
  \label{tab:bfs_reattachment}
  \small
  \begin{tabular}{lccc}
    \toprule
    Length scale & $x_r/H$ & $(x_r - x_{r,\text{exp}})/H$ & $\delta_{x_r}$ \\
    \midrule
    Experiment              & 6.67  & --              & --              \\
    $\Delta_{\max}$         & 12.50 & +5.83  & 87.4\% \\
    $\Delta_{\text{swirl}}$ & 11.40 & +4.73  & 71.0\% \\
    $\Delta_{\lambda_2}$    & 10.79 & +4.12  & 61.8\% \\
    $\Delta_{\omega}$       & 10.07 & +3.40  & 51.0\% \\
    $\Delta_{\text{SLA}}$   & 7.54  & +0.87  & 13.0\% \\
    $\Delta_{R}$            & 7.13  & +0.46  & 6.9\%  \\
    \bottomrule
  \end{tabular}
\par\end{table}

\begin{figure}[hbt!]
  \centering
  \includegraphics[width=0.8\textwidth]{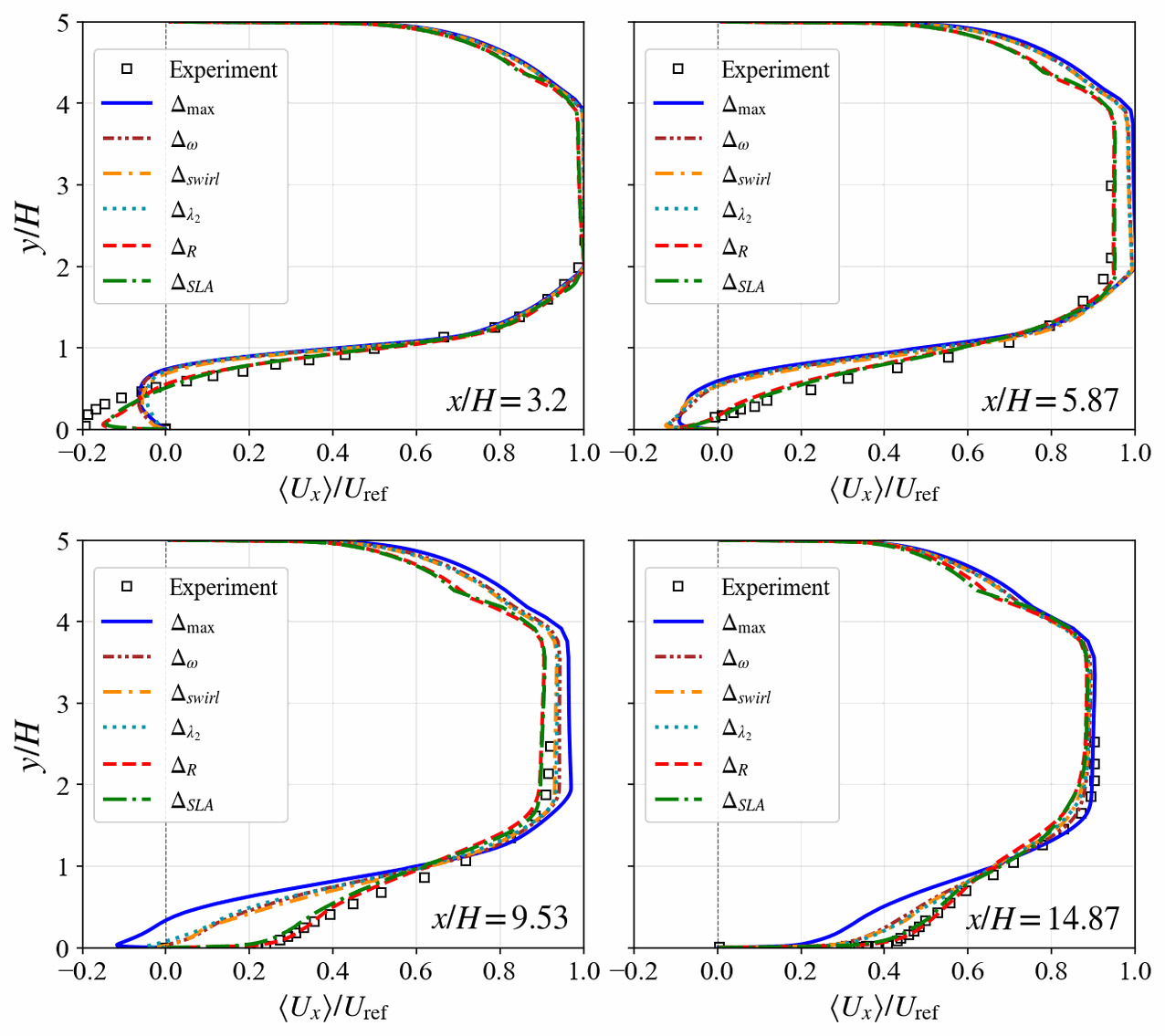}
  \caption{Dimensionless velocity profiles ${\langle U_x\rangle/U_{ref}}$ distributions obtained using different length scales in backward-facing step.}
  \label{fig:bfs_velocity_profile}
\end{figure}

Figure~\ref{fig:bfs_velocity_profile} shows the dimensionless mean velocity profiles ${\langle U_x\rangle/U_{ref}}$ distributions obtained using different length scales in backward facing step. At $x/H=3.2$ and $x/H=5.87$, $\Delta_{\text{max}}$, $\Delta_{\omega}$, $\Delta_{\text{swirl}}$, and $\Delta_{\lambda_2}$ yield similar velocity profiles, while $\Delta_{\mathrm{R}}$ and $\Delta_{\text{SLA}}$ show comparable results. At $x/H=9.53$ and $x/H=14.87$, the accuracy improves in the order of $\Delta_{\text{max}}$, $\Delta_{\text{swirl}}$, $\Delta_{\lambda_2}$, $\Delta_{\omega}$, $\Delta_{\text{SLA}}$, and $\Delta_{\mathrm{R}}$.

\begin{figure}[hbt!]
  \centering
  \includegraphics[width=0.8\textwidth]{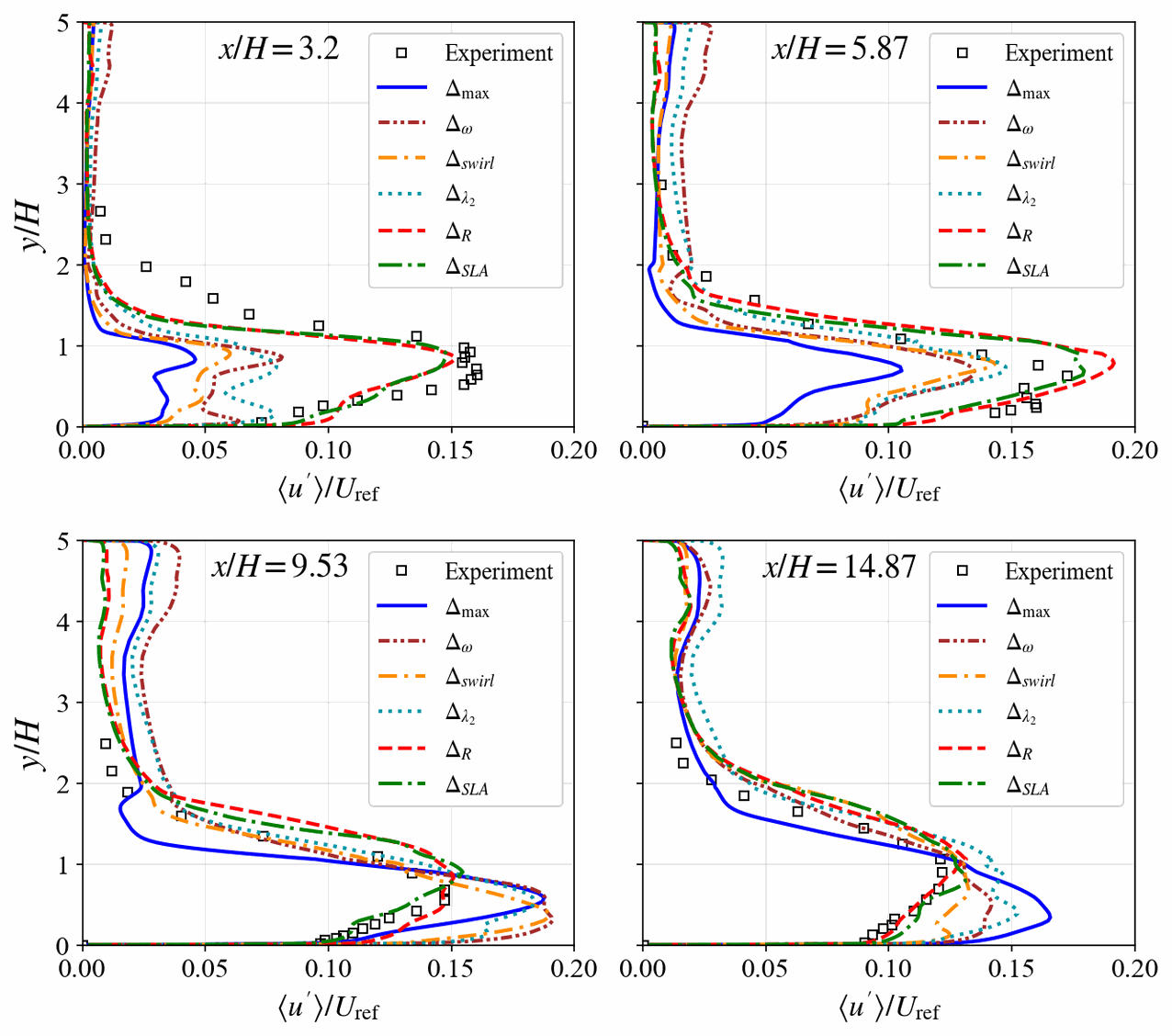}
  \caption{Dimensionless streamwise rms velocity fluctuation $u_{\mathrm{rms}}/U_{ref}$ distributions obtained using different length scales in backward-facing step.}
  \label{fig:bfs_Reynolds_Stress_u}
\end{figure}

Figure~\ref{fig:bfs_Reynolds_Stress_u} shows the dimensionless streamwise root-mean-square (rms) velocity fluctuation $u_{\mathrm{rms}}/U_{ref}$, where $u_{\mathrm{rms}}=\sqrt{\langle u'u' \rangle}$, distributions obtained using different length scales. At $x/H=3.2$, which lies in the early stage of the separated shear layer, $\Delta_{\text{max}}$ and $\Delta_{\text{swirl}}$ severely under-predict the streamwise velocity fluctuations, with peak values of approximately $0.05$, compared with approximately $0.16$ in the experiment. $\Delta_{\omega}$ and $\Delta_{\lambda_2}$ partially recover the fluctuation level but still under-predict the peak. In contrast, $\Delta_{R}$ and $\Delta_{\text{SLA}}$ reproduce both the magnitude and the wall-normal location of the measured peak. This indicates that, with $\Delta_{R}$ and $\Delta_{\text{SLA}}$, resolved turbulent fluctuations develop rapidly downstream of the step edge, whereas the other length scales delay the K--H instability and the transition from modeled to resolved turbulence. At $x/H=5.87$, the fluctuations predicted by $\Delta_{\text{max}}$ and $\Delta_{\text{swirl}}$ remain too low in the lower part of the shear layer ($y/H<0.6$), while $\Delta_{R}$ and $\Delta_{\text{SLA}}$ remain closest to the experimental profile. Further downstream, at $x/H=9.53$ and $x/H=14.87$, the trend reverses for $\Delta_{\text{max}}$, $\Delta_{\omega}$, $\Delta_{\text{swirl}}$, and $\Delta_{\lambda_2}$, which over-predict the peak fluctuations. This overshoot is a typical consequence of the delayed K--H instability, in which the resolved fluctuations develop late and then grow excessively. $\Delta_{R}$ and $\Delta_{\text{SLA}}$ avoid this overshoot and agree well with the measured peak values. Therefore, $\Delta_{R}$ and $\Delta_{\text{SLA}}$ provide the best overall predictions of the streamwise velocity fluctuations among all considered length scales, which is consistent with the reduced eddy viscosity in the separated shear layer (figure~\ref{fig:bfs_eddy_viscosity}) and the improved $C_f$ and mean-velocity predictions (figures~\ref{fig:bfs_Cf} and \ref{fig:bfs_velocity_profile}).

\subsubsection{Free shear layer}
\label{sec:Free shear layer}

\begin{figure}[hbt!]
  \centering  \includegraphics[width=0.6\textwidth]{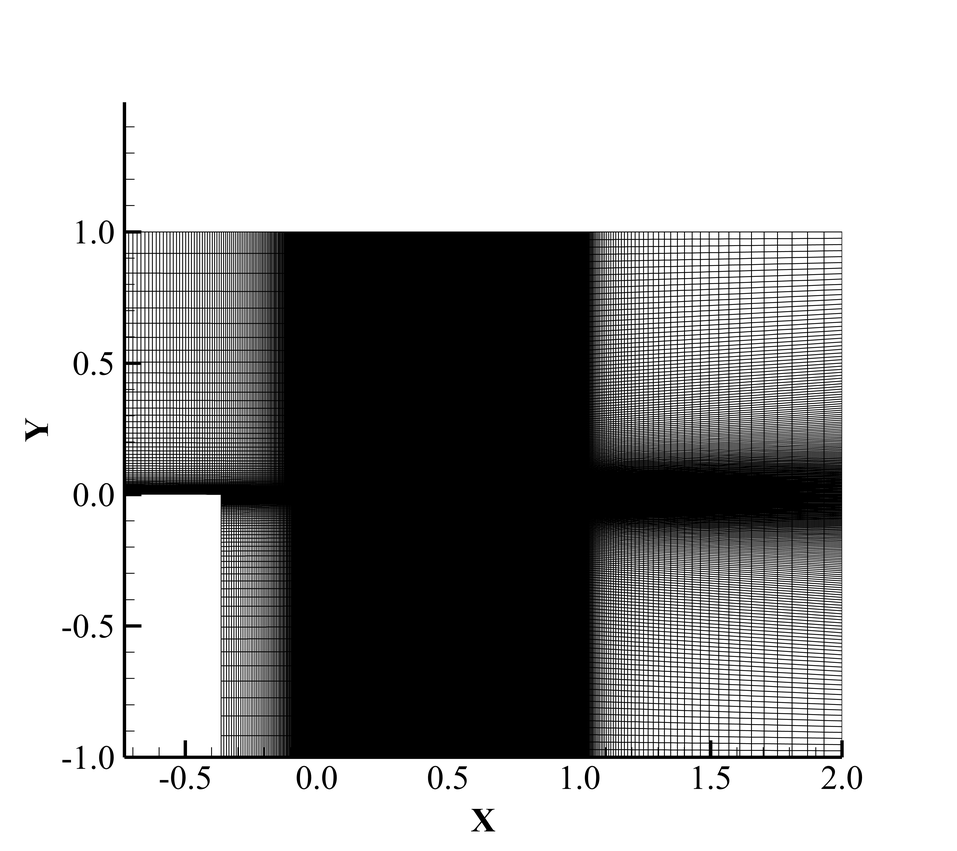}
  \caption{Computational domain and grids for free shear layer.}
  \label{fig:fsl_grid}
\end{figure}

\begin{figure}[hbt!]
  \centering  \includegraphics[width=0.6\textwidth]{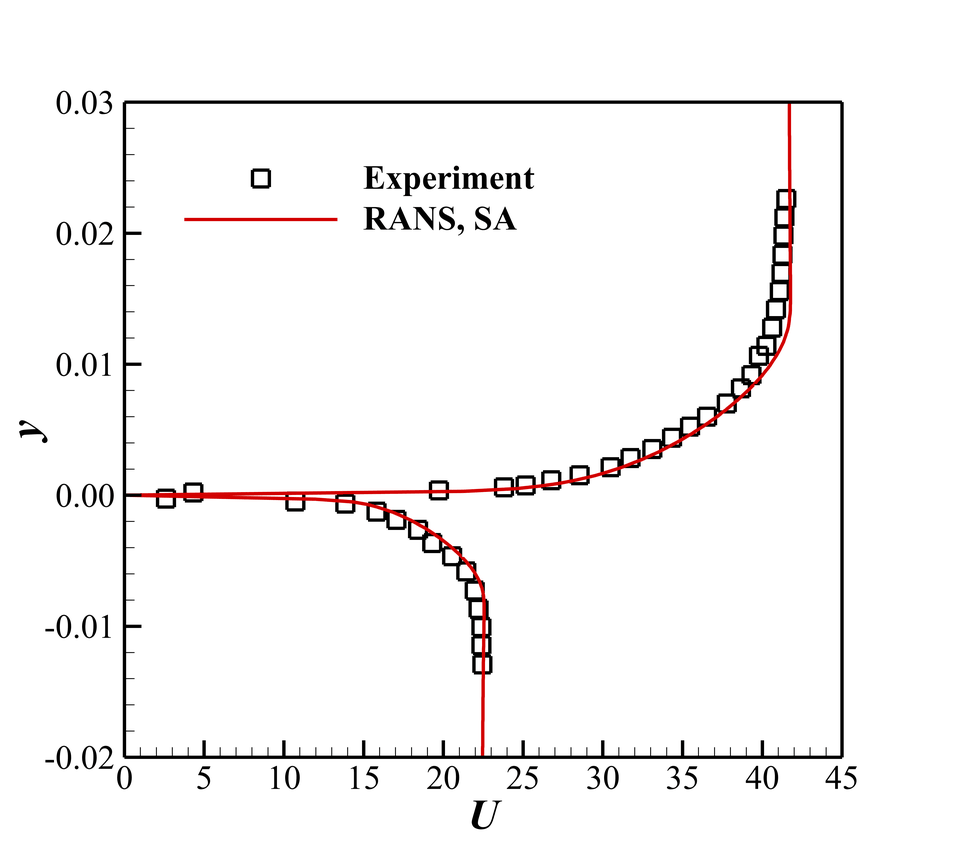}
  \caption{Streamwise velocity profile at 1 mm
downstream of the trailing edge in free shear layer.}
  \label{fig:fsl_velocity_profile}
\end{figure}

  The freestream velocities of the upper and lower layers are $U_1$=41.54m/s and $U_2$=22.40m/s, respectively. The free shear layer develops over a zero-thickness flat-plate model and is generated by the interaction of the boundary layers formed on both sides of the plate.  Figure~\ref{fig:fsl_grid} shows the computational domain of the free shear layer. The grid is uniformly distributed in the streamwise direction, with dx=dz in the region $x/L=0 \text{--} 1$. The flat-plate length on the high-velocity side is set to 730 mm, whereas that on the low-velocity side is 360 mm. The total number of grid cells is approximately 18 million. To obtain a satisfactory boundary-layer profile at the trailing edge of the flat plate (x=0.0m), the inlet section is extended upstream. As shown in figure~\ref{fig:fsl_velocity_profile}, this configuration ensures that the streamwise velocity profile at 1 mm downstream of the trailing edge agrees well with the experimental data reported by \citet{delville1995decomposition}.

\begin{figure}[hbt!]
  \centering
  \begin{subfigure}[b]{0.45\textwidth}
    \includegraphics[width=\textwidth]{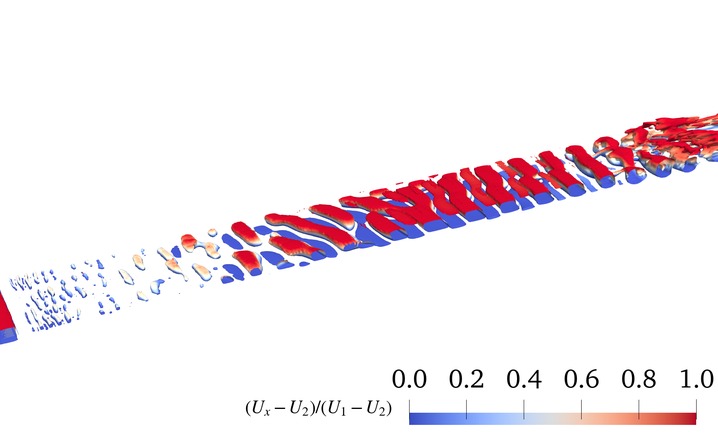}
    \caption{$\Delta_{\max}$}
  \end{subfigure}\hfill
  \begin{subfigure}[b]{0.45\textwidth}
    \includegraphics[width=\textwidth]{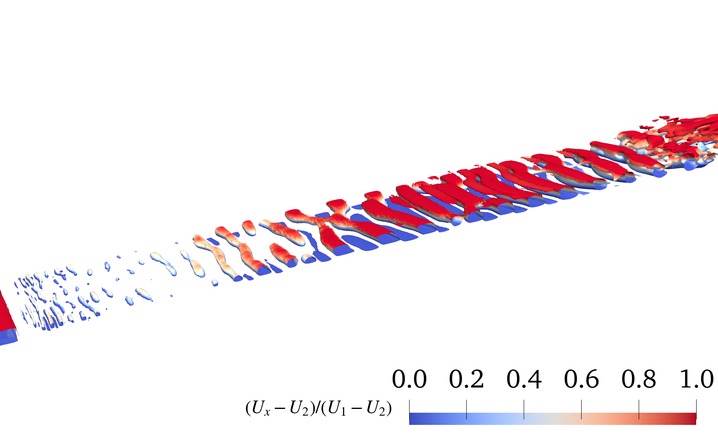}
    \caption{$\Delta_{\mathrm{\omega}}$}
  \end{subfigure}
  \begin{subfigure}[c]{0.45\textwidth}
    \includegraphics[width=\textwidth]{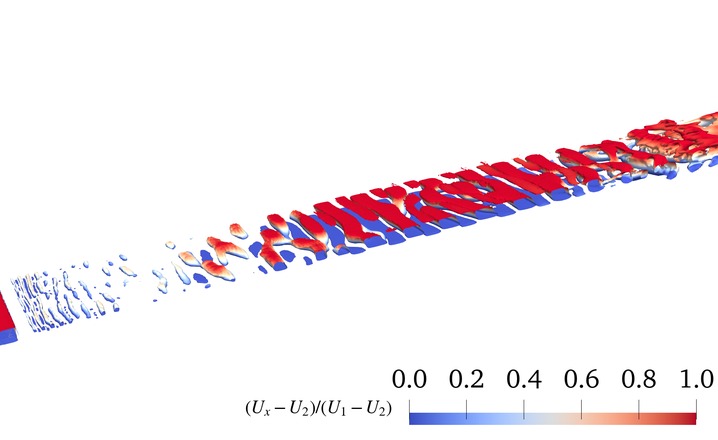}
    \caption{$\Delta_{\lambda_2}$}
  \end{subfigure}
  \begin{subfigure}[d]{0.45\textwidth}
    \includegraphics[width=\textwidth]{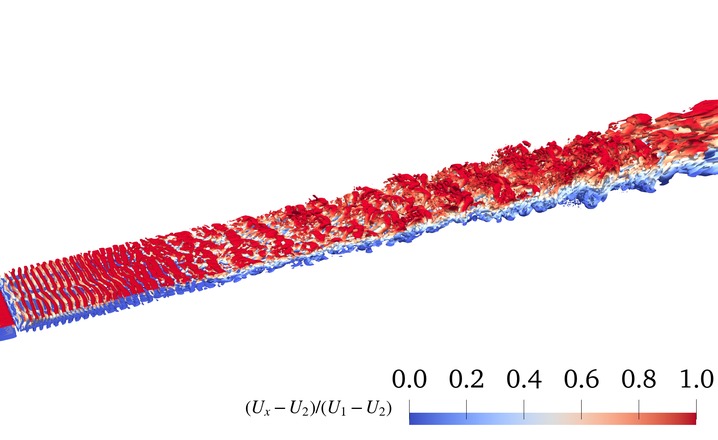}
    \caption{$\Delta_{\mathrm{R}}$}
  \end{subfigure}
  \begin{subfigure}[d]{0.45\textwidth}
    \includegraphics[width=\textwidth]{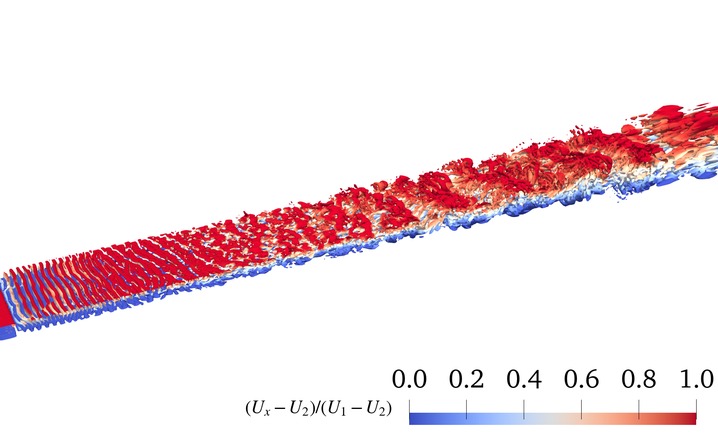}
    \caption{$\Delta_{\mathrm{SLA}}$}
  \end{subfigure}
  \caption{Different vortex evolution processes using different length scales in free shear layer. Isosurfaces of the dimensionless Q criterion, $Q(L/U_1)^2=0.58$, colored by $(U_x-U_2)/(U_1-U_2)$.}
  \label{fig:fsl_Q}
\end{figure}

Figure~\ref{fig:fsl_Q} illustrates the different vortex evolution processes obtained using different length scales. The vortical structures are visualized by isosurfaces of the dimensionless Q criterion, $Q(L/U_1)^2=0.58$ (i.e., $Q=1000$~s$^{-2}$ with $L=1$~m), colored by the normalized streamwise velocity $(U_x-U_2)/(U_1-U_2)$. The same isosurface value is used in all five panels. Only the candidates that pass the BFS case are shown, together with the baseline $\Delta_{\max}$. $\Delta_{\text{swirl}}$ is discarded after the BFS case and is not shown. The $\Delta_{\text{max}}$, $\Delta_{\omega}$, and $\Delta_{\lambda_2}$ length scales exhibit excessive numerical dissipation and suppress the development of the Kelvin--Helmholtz (K--H) instability, whereas the $\Delta_{R}$ length scale provides an improved prediction of the K--H instability. As shown in figure~\ref{fig:fsl_Q} (d) and (e), spanwise vortices, K--H instabilities, and hairpin vortices can be clearly observed downstream of the flat-plate trailing edge.

\begin{figure}[hbt!]
  \centering
  \includegraphics[width=0.9\textwidth]{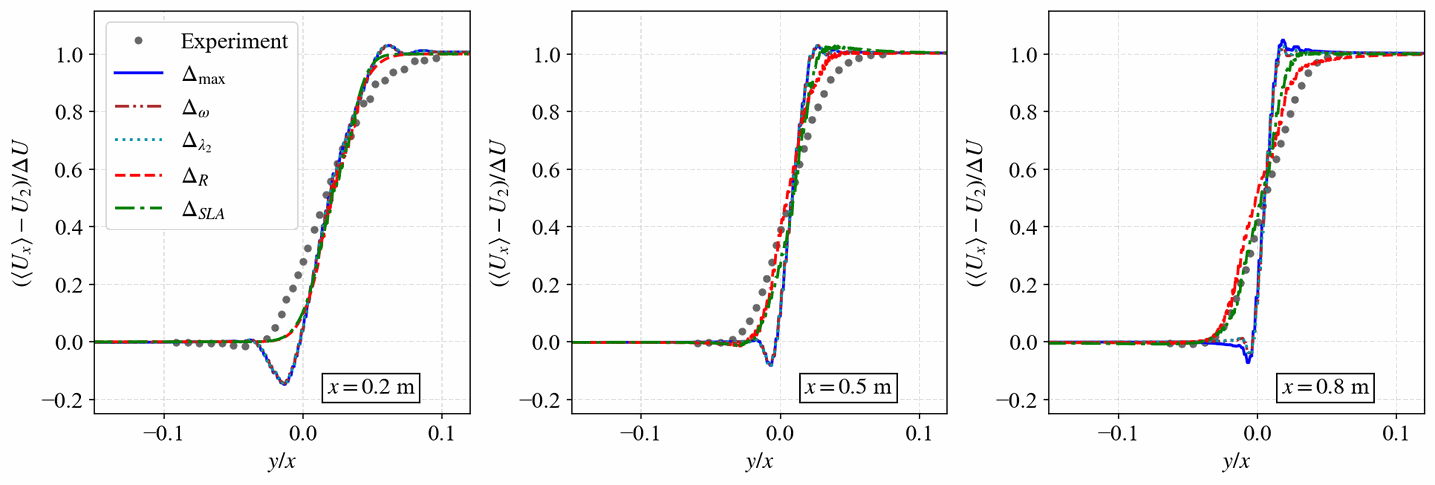}
  \caption{Mean streamwise velocity profiles obtained using the different length scales in free shear layer.}
  \label{fig:fsl_velocity_profiles}
\end{figure}

Figure~\ref{fig:fsl_velocity_profiles} compares the mean streamwise velocity profiles obtained using different length scales. In the vertical axis, $\Delta U$ denotes $U_1 - U_2$. Statistical averaging starts after the flow at $U_2=22.40$ m/s passes through 7 computational domains. The subsequent 5 flow-through times are used to obtain the time-averaged flow field. The profiles obtained with $\Delta_{\text{max}}$, $\Delta_{\omega}$, and $\Delta_{\lambda_2}$ are nearly identical at all three streamwise locations, $x=0.2$ m, $x=0.5$ m, and $x=0.8$ m. In contrast, $\Delta_{R}$ and $\Delta_{\text{SLA}}$ provide significantly improved predictions of the velocity profiles. At $x=0.2$~m and $x=0.5$~m, the three length scales $\Delta_{\max}$, $\Delta_{\omega}$, and $\Delta_{\lambda_2}$ retain a velocity deficit below the low-speed freestream velocity on the low-speed side of the layer, which is a remnant of the wake of the splitter plate and indicates delayed mixing. $\Delta_{R}$ and $\Delta_{\text{SLA}}$ eliminate this deficit and, at $x=0.8$~m, reproduce the spreading of the layer more closely. The profiles of $\Delta_{R}$ and $\Delta_{\text{SLA}}$ are similar at all three locations, so both candidates are retained for the final case. Quantitatively, $\Delta_R$ and $\Delta_{\text{SLA}}$ give $\varepsilon_U = 0.161$ and $0.163$, respectively, compared with $0.259$ for $\Delta_{\omega}$ and $0.262$ for $\Delta_{\lambda_2}$ (table~\ref{tab:error_metrics}), i.e., a reduction of the error by about $38\%$.

\subsubsection{GLC305 with 944 ice shape}
\label{sec:GLC305 with 944 ice shape}
\begin{figure}[hbt!]
  \centering  \includegraphics[width=1.0\textwidth]{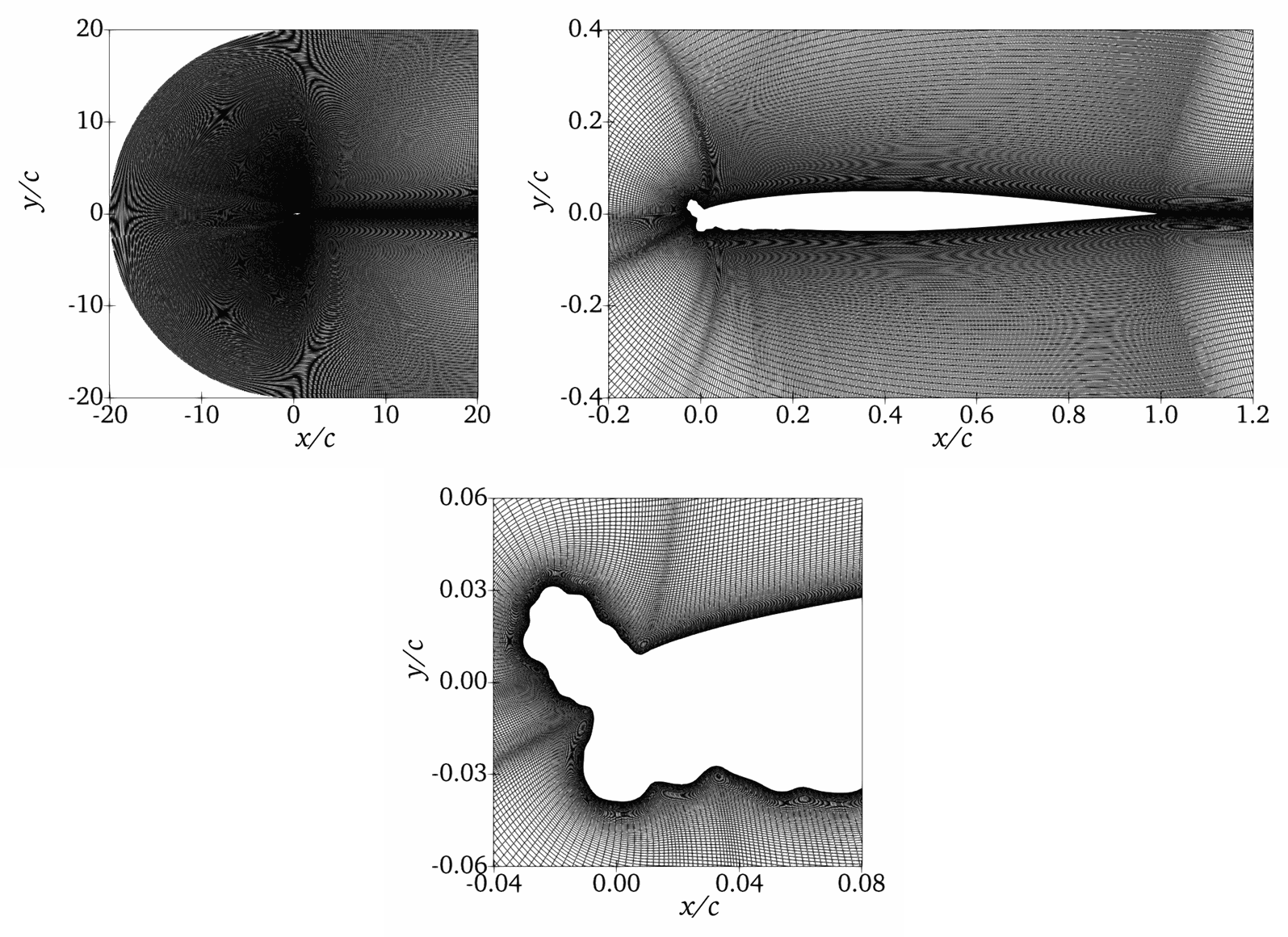}
  \caption{Computational domain and grids for GLC305 with 944 ice shape.}
  \label{fig:GLC305_grid}
\end{figure}

The final case is a GLC305 airfoil with the 944 ice shape. It is used to assess whether the proposed length scales are applicable to high-Reynolds-number separated flows. The iced GLC305 airfoil case has a glaze ice 944 at the leading edge. Ice shape 944 has ice horns on both the upper and lower surfaces and has a 
relative ice height of $h/c = 3.68\%$. The flow conditions are $\mathit{Ma} = 0.12$, 
$\mathit{Re} = 3.5 \times 10^6$, and $AoA = 6^\circ$. 
Relevant experimental results can be found in \citet{broeren2006flowfield}. Two grid sets with approximately 10 million and 27 million cells are used for the IDDES simulations. Figure~\ref{fig:GLC305_grid} shows the computational domain and grids for GLC305 with 944 ice shape. A C-type grid is employed, with dimensions of 1265$\times$241$\times$35. The first layer grid height corresponds to $y^+\approx0.96$. This grid corresponds to the coarser grid set with approximately 10 million cells.

\begin{figure}[hbt!]
\centering
\begin{subfigure}[b]{0.45\textwidth}
    \centering
    \includegraphics[width=\textwidth]{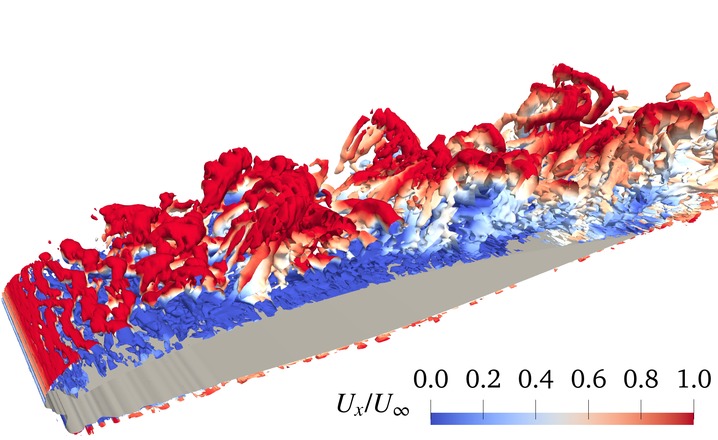}
    \caption{$\Delta_{\max}$}
\end{subfigure}
\hfill %
\begin{subfigure}[b]{0.45\textwidth}
    \centering
    \includegraphics[width=\textwidth]{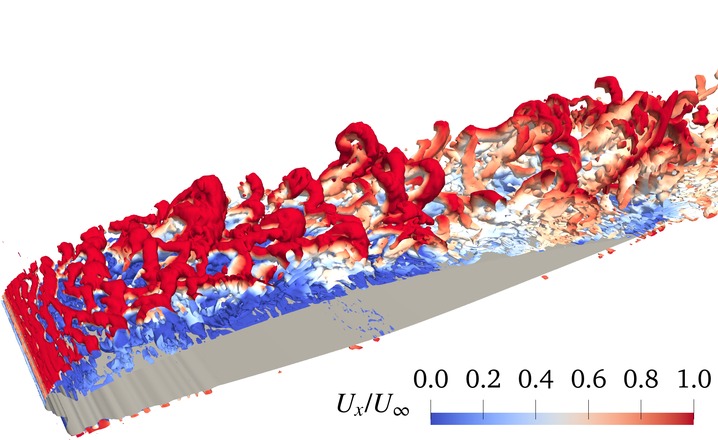}
    \caption{$\Delta_{\mathrm{R}}$}
\end{subfigure}

\vspace{1em} %

\begin{subfigure}[b]{0.45\textwidth}
    \centering
    \includegraphics[width=\textwidth]{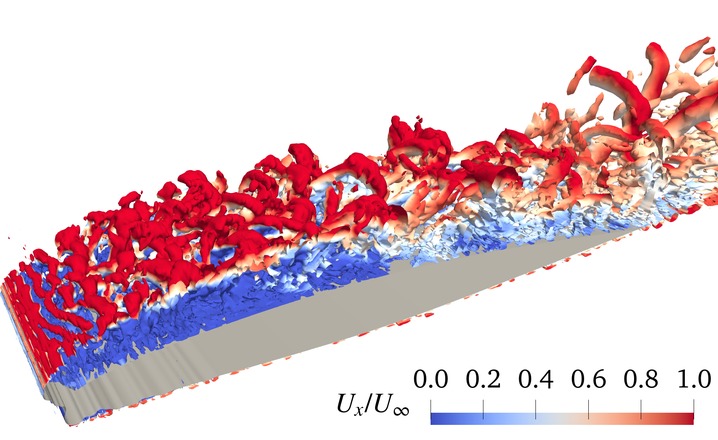}
    \caption{$\Delta_{\mathrm{SLA}}$}
\end{subfigure}

\caption{Isosurfaces of the dimensionless Q criterion, $Q(c/U_\infty)^2=50$, colored by the dimensionless streamwise velocity ($U/U_\infty$) for GLC305 with 944 ice shape ($Ma = 0.12$, $Re = 3.5 \times 10^6$, and $\text{AoA} = 6^\circ$).}
\label{fig:GLC305_Q}
\end{figure}

Figure~\ref{fig:GLC305_Q} shows the isosurfaces of the dimensionless Q criterion, $Q(c/U_{\infty})^2=50$, colored by the dimensionless streamwise velocity ($U/U_{\infty}$). The same isosurface value is used in all three panels. Only $\Delta_R$ and $\Delta_{\text{SLA}}$ enter this case. They are shown together with the baseline $\Delta_{\max}$. The flow field exhibits two-dimensional vortices, Kelvin--Helmholtz (K--H) instabilities, and hairpin vortices. Due to the leading-edge ice shape, a separation bubble forms immediately downstream of the ice, with a fixed separation point.

\begin{figure}[hbt!]
  \centering
  \includegraphics[width=0.6\textwidth]{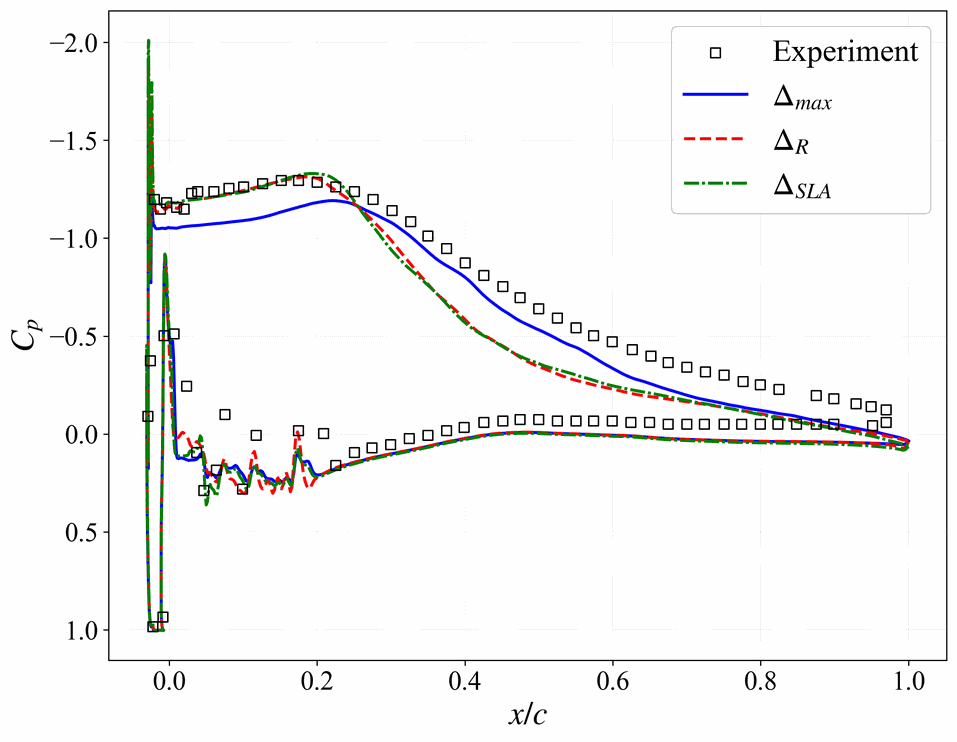}
  \caption{Comparison of pressure coefficient $C_p$ distributions obtained using different length scales for GLC305 with 944 ice shape ($\mathit{Ma} = 0.12$, 
$\mathit{Re} = 3.5 \times 10^6$, and $AoA = 6^\circ$).}
  \label{fig:GLC305_Cp}
\end{figure}

Figure~\ref{fig:GLC305_Cp} shows the Comparison of pressure coefficient $C_p$ distributions obtained using different length scales. The length scale $\Delta_{\mathrm{max}}$ predicts the lowest suction plateau. In contrast, $\Delta_{\mathrm{R}}$ provides an accurate prediction of the suction plateau. $\Delta_{\text{SLA}}$ yields a $C_p$ distribution almost identical to that of $\Delta_{R}$ over the suction plateau ($x/c<0.25$). Downstream of the plateau ($x/c>0.3$), all length scales under-predict the suction during pressure recovery, and $\Delta_{R}$ and $\Delta_{\text{SLA}}$ show a larger deviation from the experiment than $\Delta_{\max}$ in this region. Quantitatively, $\Delta_R$ gives $\varepsilon_{C_p}=0.359$ and $\Delta_{\text{SLA}}$ gives $\varepsilon_{C_p}=0.365$ (table~\ref{tab:error_metrics}). $\Delta_R$ is therefore ranked first and returned as the final output. However, the difference between the two errors is only about $2\%$, which is comparable to the uncertainty expected from the finite averaging time. The two length scales should thus be regarded as performing equally well in this case. The final selection of $\Delta_R$ over $\Delta_{\text{SLA}}$ mainly reflects that $\Delta_R$ achieves this accuracy without the empirical kinematic sensor $F_{KH}$ and its four calibration constants. The earlier pressure recovery predicted by $\Delta_R$ and $\Delta_{\text{SLA}}$ indicates a shorter separation bubble than in the experiment. Several factors may contribute to this discrepancy. First, the spanwise extent of the domain is $0.3c$, which may constrain the large-scale three-dimensional structures that develop near reattachment. Second, the results are obtained on the coarser grid with 35 cells in the spanwise direction, and the resolution of the reattaching shear layer may be insufficient. Third, the lower-surface $C_p$ is also offset from the measured values over most of the chord for all length scales, which suggests a difference in circulation. This difference may be associated with wind-tunnel wall effects that are not represented in the free-air simulations. Because $\Delta_{\max}$ predicts a lower suction plateau but a later pressure recovery, its smaller error in the recovery region is likely the result of error cancellation rather than a better representation of the shear-layer dynamics.

\subsection{Physical Interpretation of different Length Scales}

This section interprets the different length scales from two complementary perspectives. The first is the angular-momentum-integral (AMI) decomposition \citep{elnahhas2022enhancement,nair2026integral}, which provides a theoretical interpretation. The second is vortex-axis identification, which offers a more intuitive physical understanding.


\subsubsection{Perspective from angular-momentum-integral (AMI) decomposition}
\label{sec:perspective_cf}
The first analysis focuses on $C_f$, using the backward-facing step to examine the behavior of different length scales.
 \citet{nair2026integral}
recently developed angular-momentum-integral (AMI) decomposition of
\citet{elnahhas2022enhancement} into a mechanism-resolved error diagnostic for turbulence models. Since
the mean BFS flow is statistically two-dimensional and wall-bounded, the skin-friction
coefficient $C_f \equiv \tau_w/(\tfrac12\rho U_{io}^2)$ along the lower wall splits exactly into

\begin{equation}
  \frac{C_f}{2} \;=\; \underbrace{\frac{1}{Re_{\ell}}}_{\text{viscous}}
  \;+\; \underbrace{\int_0^{\infty} \frac{\tau_{app}(x,y)}{\rho\,U_{io}^2\,\ell}\,\mathrm{d}y}_{\text{turbulent torque}}
  \;+\; \mathcal{F}_{\text{geo}}(x),
  \label{eq:ami}
\end{equation}

where $U_{io}$ is the local irrotational reference velocity, $\ell$ is the boundary-layer
moment arm, $Re_{\ell}=\rho U_{io}\ell/\mu$, and $\mathcal{F}_{\text{geo}}(x)$ collects the
mean-flux, pressure-gradient and boundary-layer-departure contributions. The budget remains
well posed inside the BFS recirculation bubble \citep{kianfar2025moment}. Since $\mathcal{F}_{\text{geo}}$
depends only on the mean velocity, the subgrid length scale enters equation~\eqref{eq:ami}
explicitly only through the turbulent-torque term. In an IDDES solution, the shear stress in
this term is an apparent stress containing both resolved and modelled parts,

\begin{equation}
  \tau_{app}(x,y) \;=\; -\rho\,\overline{u'v'} \;+\; \mu_t\left(\frac{\partial \bar U}{\partial y}+\frac{\partial \bar V}{\partial x}\right),
  \label{eq:tau_app}
\end{equation}
where $\overline{u'v'}$ is the resolved Reynolds shear stress and $\mu_t$ is the eddy
viscosity obtained from equation~\eqref{eq:nu-transport}.

\begin{figure}[hbt!]
  \centering
  \includegraphics[width=0.8\textwidth]{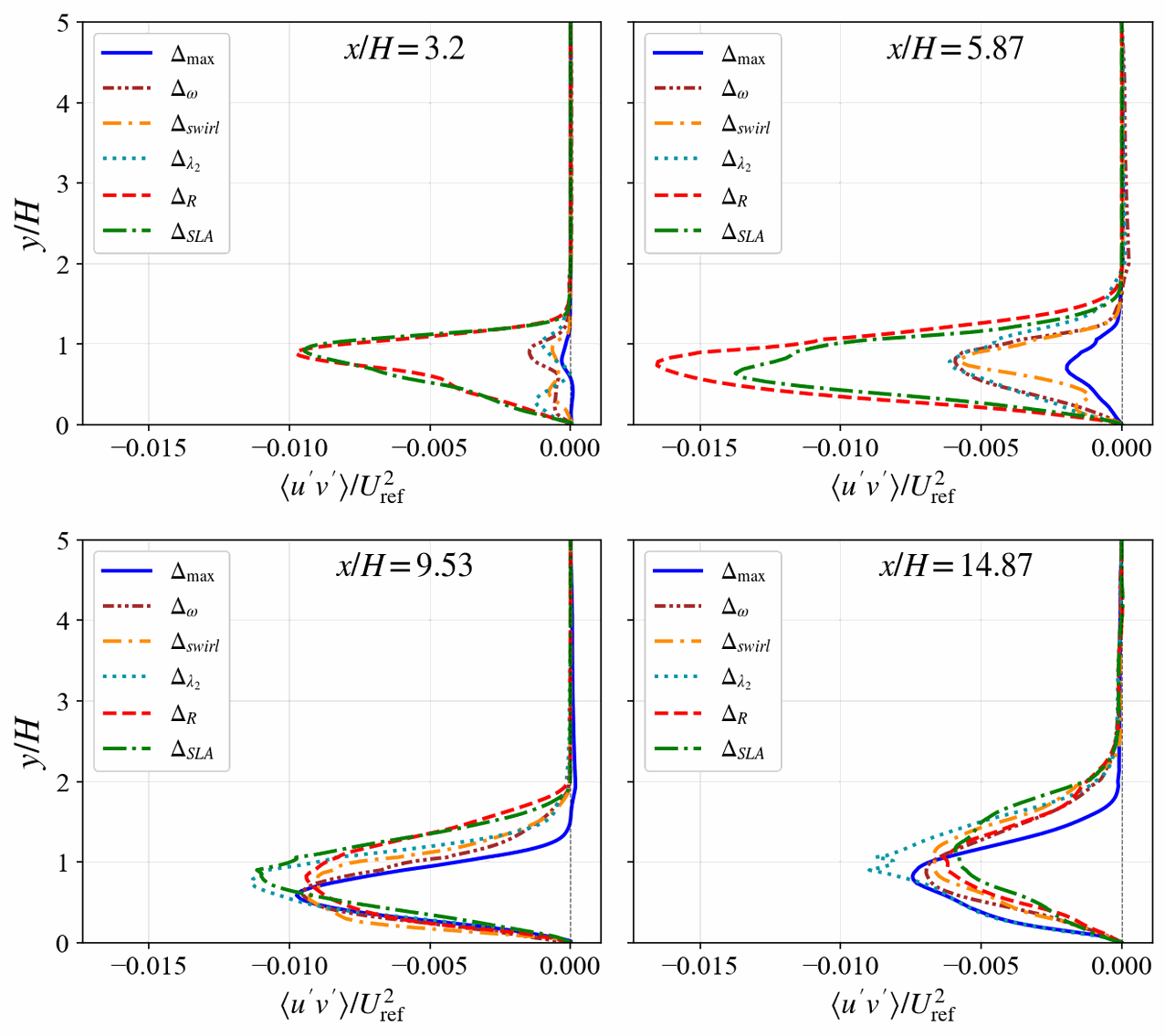}
  \caption{Dimensionless Reynolds stress component ${<u'v'>/U_{ref}^2}$ distributions obtained using different length scales in backward-facing step.}
  \label{fig:bfs_Reynolds_Stress_uv}
\end{figure}

The length scale controls $\tau_{app}$ through the destruction term in equation~\eqref{eq:nu-transport}, which scales as $(\tilde\nu/l_{\text{IDDES}})^2$. A larger $l_{\text{IDDES}}$ weakens the destruction term and slows the decay of $\mu_t$. In the anisotropic cells of the shear layer separating from the step edge, $\Delta_{\max}$ selects the largest cell dimension. This increases $l_{\text{LES}}$ and allows $\mu_t$ to persist over a longer streamwise distance downstream of separation, as shown in figure~\ref{fig:bfs_eddy_viscosity}. The resulting excess viscosity damps the small perturbations that drive Kelvin--Helmholtz instability, thereby delaying the formation of the rollers observed with $\Delta_{\max}$ in figure~\ref{fig:bfs_Q}. Within this grey-area region, neither contribution in equation~\eqref{eq:tau_app} is adequately represented. The modelled contribution mainly diffuses the smooth mean shear, while the resolved contribution $\overline{u'v'}$ is suppressed by the excess viscosity. Consequently, $\tau_{app}$ is under-predicted, leading to an under-prediction of the turbulent-torque term in equation~\eqref{eq:ami}. Figure~\ref{fig:bfs_Reynolds_Stress_uv} compares the resolved Reynolds shear stress $\langle u'v'\rangle/U_{ref}^2$ obtained with different length scales, providing a direct measure of the resolved contribution to $\tau_{app}$.

$\Delta_R$ corrects the grid-scale deficiency at its source. By measuring the cell dimension normal to the local vortex axis, rather than $\max(\Delta x,\Delta y,\Delta z)$, $\Delta_R$ is smaller than $\Delta_{\max}$ in the pencil- and book-type cells within the separating shear layer. The resulting reduction in eddy viscosity (figure~\ref{fig:bfs_eddy_viscosity}) promotes earlier development of the Kelvin--Helmholtz instability. This shortens the streamwise region with under-predicted $\tau_{app}$ and leads to a smaller separation region (figure~\ref{fig:bfs_Q}). In contrast, the attached boundary layer upstream of the step remains protected by the RANS shielding function $\tilde f_d$. Although accurate $C_f$ predictions may result from error cancellation, an error in turbulent torque can be offset by an opposite error in $\mathcal{F}_{\text{geo}}$. Such cancellation can fail in adverse-pressure-gradient and reattachment regions, both present downstream of the BFS step.
 Therefore, the under-predicted turbulent torque with $\Delta_{\max}$ directly leads to the largest $C_f$ discrepancy in figure~\ref{fig:bfs_Cf}. By correcting the turbulent-torque contribution itself, rather than relying on error compensation, $\Delta_R$ achieves the best overall agreement with the measured $C_f$.

\subsubsection{Perspective from vortex-axis identification}
\label{sec:Perspective from vortex-axis identification}

\begin{figure}[hbt!]
  \centering
  \includegraphics[width=0.5\textwidth]{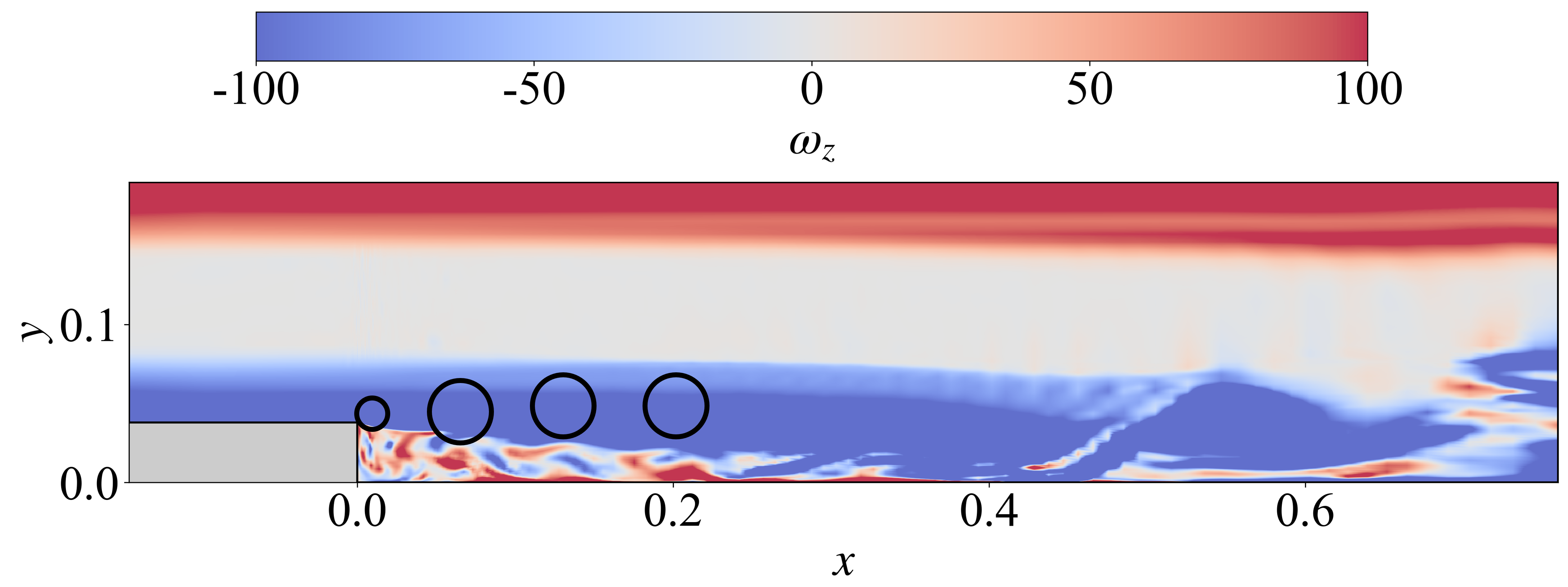}
  \caption{Separating shear layer identified by the spanwise vorticity component $\omega_z$ for backward facing step.}
  \label{fig:bfs_separating_shear_layer}
\end{figure}

The differences are further analyzed from the perspective of vortex-axis identification. An open-source multi-agent framework PhysMiner \citep{chen2026physminer} is employed to perform triple decomposition and examine the underlying flow physics. Since PhysMiner does not currently support vortex-axis identification, the framework is extended to incorporate this capability. The backward-facing step flow computed by the length scale $\Delta_{max}$ is used as the flow-field data source for analysis. Figure~\ref{fig:bfs_separating_shear_layer} shows the separating shear layer identified by the spanwise vorticity component $\omega_z$ for backward facing step. Several circles are highlighted within the separation layer to indicate the regions examined in the following comparison.

\begin{figure}[hbt!]
  \centering

  \begin{subfigure}[b]{0.5\textwidth}
    \centering
    \includegraphics[width=\textwidth]{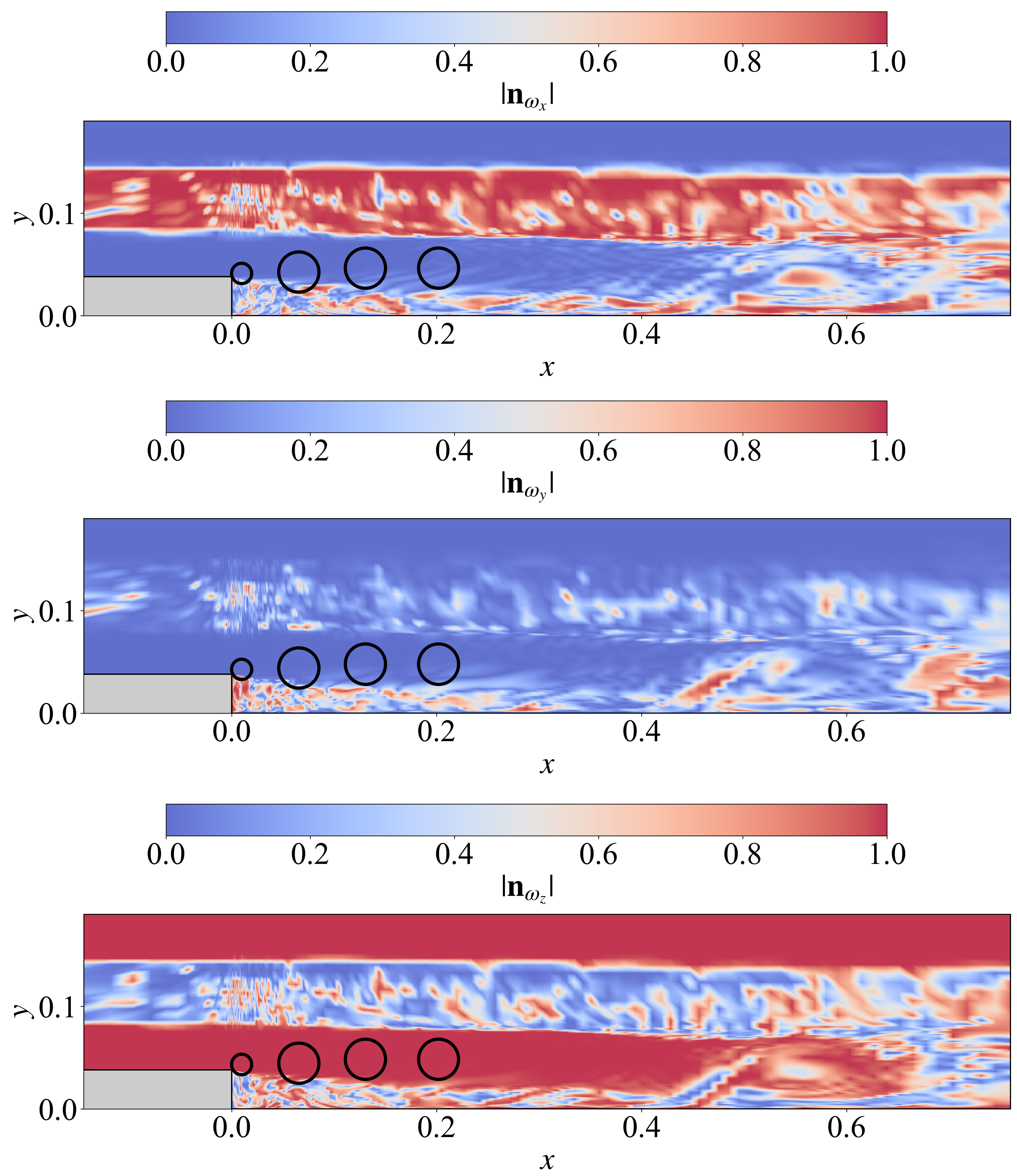}
    \caption{Vortex-axis directions identified by $\omega$}
    \label{fig:bfs_axis_a}
  \end{subfigure}%
  \hfill
  \begin{subfigure}[b]{0.5\textwidth}
    \centering
    \includegraphics[width=\textwidth]{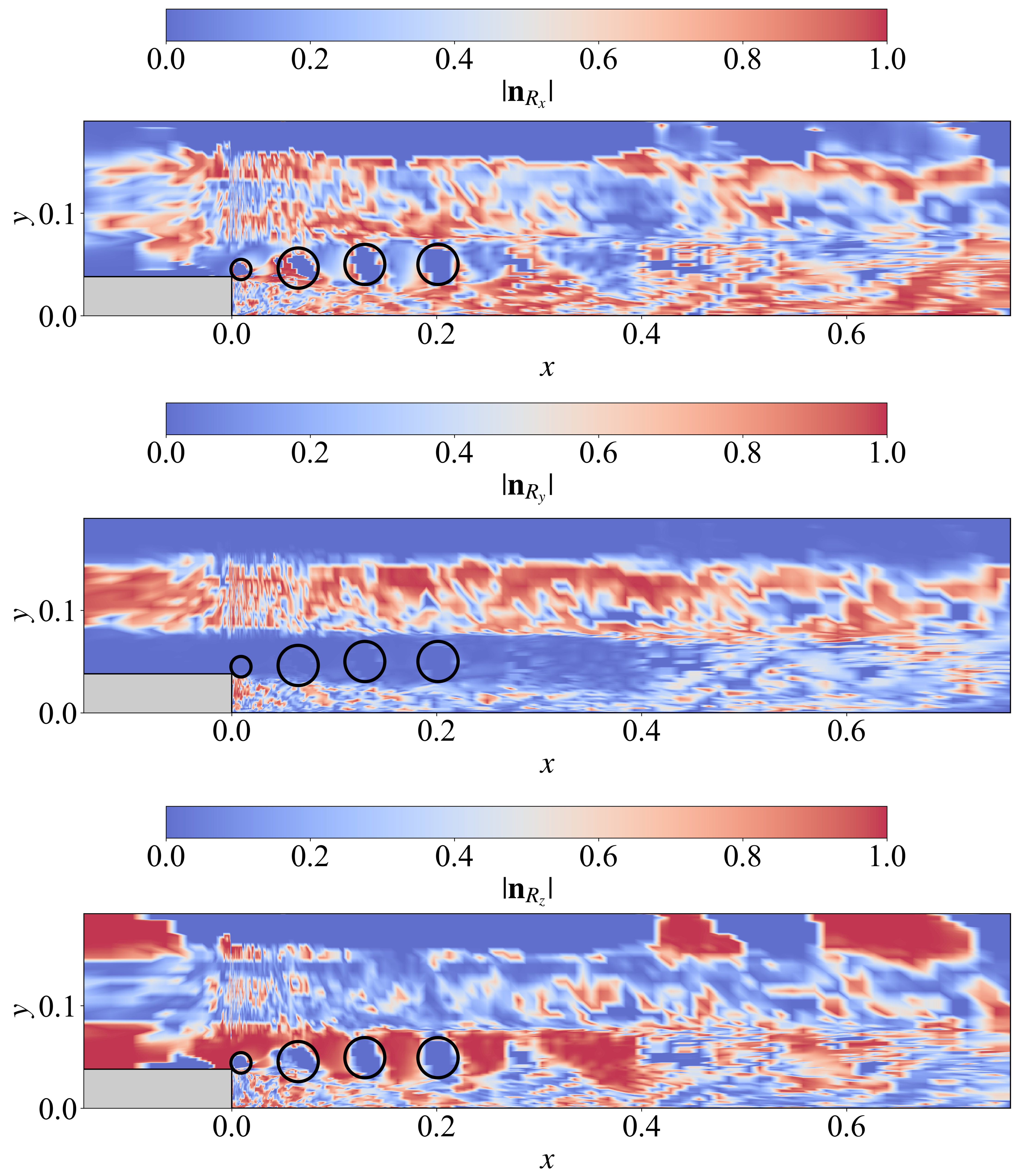}
    \caption{Vortex-axis directions identified by $R$}
    \label{fig:bfs_axis_b}
  \end{subfigure}

  \caption{Vortex-axis directions identified by $\omega$ and $R$, corresponding to the two length scales, $\Delta_{\omega}$ and $\Delta_{R}$.}
  \label{fig:bfs_vortex_axis_comparison}
\end{figure}

Figure~\ref{fig:bfs_vortex_axis_comparison} compares the vortex-axis directions identified by $\omega$ and $R$, corresponding to the two length scales, $\Delta_{\omega}$ and $\Delta_{R}$. The $R$ method identifies several circular regions with no vortex-axis direction, as these regions contain no vortex. In contrast, the $\omega$ method still identifies a mainly spanwise vortex-axis direction in these regions. This difference leads to different cell-width selections and, consequently, different eddy-viscosity generations. Specifically, $\Delta_{\omega}$ selects the cell width normal to the spanwise direction and generates eddy viscosity, whereas $\Delta_{R}$ does not select a cell width and generates no eddy viscosity. This difference allows $\Delta_{R}$ to better alleviate the delay of K--H instability than $\Delta_{\omega}$.

It should be noted that, in cells where no vortex axis is identified, $\Delta_R$ is determined by the fallback treatment of Proposal 3 (Appendix~\ref{app:proposals}). In these cells, the Liutex vector vanishes, $\mathbf{n}_R=\mathbf{0}$, and $\Delta_R=0$. The outer length scale is then given by the lower bound of equation~\eqref{eq:cdes_amd_lim}, $\Delta_{\text{outer},lim}=C_{lim}V^{1/3}=0.15\,V^{1/3}$, which is roughly an order of magnitude smaller than $\Delta_{\max}$. $\Delta_R$ suppresses the subgrid viscosity in the non-vortical, quasi-two-dimensional region of the initial shear layer. In this respect, $\Delta_R$ acts in a similar way to $\Delta_{\text{SLA}}$, in which $F_{KH}$ reduces the length scale to $F_{KH}^{\min}\Delta_{\omega}=0.1\,\Delta_{\omega}$ in the same region. This is consistent with the similar results of the two length scales in all cases.

The same reasoning explains the ranking of the other candidates in table~\ref{tab:error_metrics}. Consider the initial region of the shear layer, where the resolved flow is close to a simple shear $\partial U/\partial y$. For $\Delta_{\text{swirl}}$, the eigenvalues of $\nabla\mathbf{u}$ are real, and the fallback of Proposal~2 assigns $C_\delta \max_f(2|\mathbf{r}_f|)$, which equals $C_\delta\Delta_{\max}$ on hexahedral cells. $\Delta_{\text{swirl}}$ therefore recovers $\Delta_{\max}$ in exactly the region where a reduction of the length scale is needed, which explains why its results are close to those of $\Delta_{\max}$. Although $\Delta_{\text{swirl}}$ uses the same real eigenvector as $\Delta_R$ where a swirl exists, the two candidates differ in their fallback values in non-vortical cells. For $\Delta_{\lambda_2}$, the tensor $M_{ij}=S_{ik}S_{kj}+\Omega_{ik}\Omega_{kj}$ vanishes identically for a simple shear, because $S^2$ and $\Omega^2$ cancel. The eigenvector associated with the largest eigenvalue is then determined by small perturbations and numerical noise, so $\mathbf{n}_{\lambda_2}$ is ill-defined and $\Delta_{\lambda_2}$ does not consistently reduce the length scale. For $\Delta_{\omega}$, the vorticity of a simple shear points in the spanwise direction. $\Delta_{\omega}$ therefore excludes the spanwise cell size and gives a moderate reduction of the length scale relative to $\Delta_{\max}$, but it does not reduce the length scale further in non-vortical cells. These differences are consistent with the ranking $\Delta_{\max}$, $\Delta_{\text{swirl}}$, $\Delta_{\lambda_2}$, $\Delta_{\omega}$, $\Delta_{\text{SLA}}$, and $\Delta_R$ in the BFS case (tables~\ref{tab:error_metrics} and \ref{tab:bfs_reattachment}).

\section{Conclusion}
\label{sec:conclusion}
This paper develops a multi-agent framework, DEScover, for systematically exploring candidate subgrid length scales for detached eddy simulation (DES). The framework is built on a turbulence-modeling concept that defines the local length scale from the grid width normal to the local vortex axis. Four flow cases are used for validation: a flat plate, a backward-facing step, a free shear layer, and an iced airfoil. The principal findings are summarized as follows:

(1) DEScover reduces the reliance on extensive expert knowledge in turbulence-model development. Traditional model development requires substantial expertise and time to formulate explicit mathematical expressions. DEScover leverages LLM-based reasoning to reduce this reliance and automatically generate multiple candidate formulations. This approach enables a faster and more systematic exploration of turbulence-modeling formulations.

(2) The proposed length scales are effective across a range of diverse flow cases. Among the candidate length scales, $\Delta_R$ provides the most consistent improvement across all tested cases. It maintains acceptable predictions for the flat-plate flow and alleviates the delayed onset of Kelvin--Helmholtz instability in the backward-facing step, free-shear-layer, and iced-airfoil cases.

(3) The systematic exploration of length scales reflects a progressive understanding of the underlying physics. The vorticity-based length scale provides an intuitive approach by using the cell width normal to the vortex axis. However, vorticity can be contaminated by shear effects, leading to inaccurate vortex-axis identification and requiring a correction of the direction. $\Delta_{SLA}$ successfully incorporates this directional correction. With more accurate vortex-axis identification, such a correction is no longer necessary, as demonstrated by the current $\Delta_R$.

Although DEScover has demonstrated its capability to explore new length scales for DES, several limitations remain. First, DEScover currently relies on OpenFOAM, which can be computationally expensive for systematic DES exploration. Faster solvers, such as JAX-based solvers, could substantially accelerate the exploration process. Second, DEScover is applied only to the subgrid length scale of DES in this study. Its workflow does not depend on a specific turbulence model. Further studies are needed to extend it to RANS and LES models. Such studies will show whether the framework can also support more conventional turbulence-model development and reduce the reliance on expert knowledge.

\begin{bmhead}[Acknowledgements.]
We would also like to thank Maochao Xiao for his insightful discussion in this paper. Portions of this manuscript benefited from the use of artificial intelligence tools (Claude Opus 5.5, Anthropic, medium effort setting), which were
employed primarily to improve readability, grammar, and clarity of language. All intellectual content, analysis, and
conclusions remain the sole work of the authors.
\end{bmhead}

\begin{bmhead}[Funding.]
This work was supported by the National Science Foundation under grant numbers 2415347 and 2541867.
\end{bmhead}

\begin{bmhead}[Declaration of interests.]
The authors report no conflict of interest.
\end{bmhead}

\begin{bmhead}[Data availability statement.]
The OpenFOAM implementations of the candidate length scales are openly available at \url{https://github.com/iDesign-Lab/DEScover}. The prompts, Agent Skill files and post-processed flow data that support the findings of this study are available from the corresponding author upon reasonable request.
\end{bmhead}

\begin{bmhead}[Author ORCIDs.]
J. Chen, https://orcid.org/0009-0009-3277-3474; X. I. A. Yang, https://orcid.org/0000-0003-4940-5976; P. He, https://orcid.org/0009-0009-4302-3791
\end{bmhead}

\appendix
\section{Detailed Formulations of the Rotation-Axis Proposals}
\label{app:proposals}
This appendix provides the complete formulations of Proposals 1--4 for identifying the
local vortex rotation axis $\mathbf{n}_v$, summarised in table~\ref{tab:vortex_proposals}.

Proposal 1 utilizes the direction of the vorticity vector, offering a computationally efficient approach that is widely implemented in classical CFD solvers. However, it remains sensitive to shear-dominated regions where the vorticity may not represent actual rotation motion.

\begin{equation}
\Delta_{\omega} = C_\delta \max_{f} (2| \mathbf{n}_{\omega} \times \mathbf{r}_{f} |)
\end{equation}

\begin{equation}
\mathbf{n}_{\omega} = \boldsymbol{\omega} / \|\boldsymbol{\omega}\|
\end{equation}

\begin{equation}
  \boldsymbol{\omega} =
  \left( \frac{\partial w}{\partial y} - \frac{\partial v}{\partial z} \right) \mathbf{i} +
  \left( \frac{\partial u}{\partial z} - \frac{\partial w}{\partial x} \right) \mathbf{j} +
  \left( \frac{\partial v}{\partial x} - \frac{\partial u}{\partial y} \right) \mathbf{k}
\end{equation}

Proposal 2 is based on the real eigenvector of the velocity gradient tensor, which is physically consistent with the swirling strength ($\lambda_{ci}$) criterion \citep{zhou1999mechanisms}. Compared with conventional approaches, this method exhibits improved immunity to pure shear contamination. However, a fallback strategy is still required in regions where all eigenvalues of the velocity gradient tensor are real. Specifically, when the velocity gradient tensor possesses one real eigenvalue, $\lambda_{r}$, and one pair of complex conjugate eigenvalues, $\lambda_{cr} + i\lambda_{ci}$ and $\lambda_{cr} - i\lambda_{ci}$, the local flow field is identified as vortical. In this case, the local rotation axis is defined by the real eigenvector, $\mathbf{n}_{\lambda r}$, corresponding to the real eigenvalue $\lambda_{r}$.

\begin{equation}
  (\nabla \mathbf{u} - \lambda_r \mathbf{I}) \boldsymbol{v}_{\lambda r} = \mathbf{0}
\end{equation}

\begin{equation}
\mathbf{e}_r = \boldsymbol{v}_{\lambda r} / \|\boldsymbol{v}_{\lambda r}\|
\end{equation}

\begin{equation}
\mathbf{n}_{\lambda r} = \mathbf{e}_r
\end{equation}

\begin{equation}
\Delta_{\text{swirl}} =  C_\delta \max_{f} (2| \mathbf{n}_{\lambda r} \times \mathbf{r}_{f} |)
\end{equation}

In regions where all eigenvalues of $\nabla\mathbf{u}$ are real ($\lambda_{ci}=0$), or where the computed eigenvector is numerically degenerate, no preferred direction is available, and $\Delta_{\text{swirl}}$ falls back to the direction-independent value $C_\delta \max_{f}(2|\mathbf{r}_f|)$, which equals $C_\delta\Delta_{\max}$ on hexahedral cells.

Proposal 3 introduces the Rortex (or Liutex) \citep{tian2018definitions} vector direction, which decomposes vorticity to isolate the purely rotational component. By resolving the sign ambiguity through vorticity projection, it achieves a physically unambiguous representation of the vortex axis. However, a fallback strategy is still required. When the velocity gradient tensor has one real eigenvalue, $\lambda_{r}$, and one pair of complex conjugate eigenvalues, $\lambda_{cr} + i\lambda_{ci}$ and $\lambda_{cr} - i\lambda_{ci}$, the local flow field is preliminarily identified as vortical. In this case, the real eigenvector corresponding to the real eigenvalue can be preliminarily regarded as the local rotation axis. In fact, this direction is identical to that adopted in Proposal 2. Based on this foundation, Proposal 3 further introduces two additional procedures to determine the final rotation axis. The first procedure is the sign correction shown in equation~\eqref{eq:liutex_sign}, which ensures that the direction of the rotation axis remains consistent with the vorticity vector direction. The second procedure imposes an additional criterion to exclude excessively weak rotational motions that are insufficient to form a physically meaningful vortex structure. Therefore, the identified vortex axis must also satisfy the condition given in equation~\eqref{eq:liutex_threshold}.

\begin{equation}
\mathbf{n}_R = \text{sign}(\boldsymbol{\omega} \cdot \mathbf{e}_r) \mathbf{e}_r
\label{eq:liutex_sign}
\end{equation}

\begin{equation}
(\boldsymbol{\omega} \cdot \mathbf{e}_r)^2 \geq 4 \lambda_{ci}^2
\label{eq:liutex_threshold}
\end{equation}

\begin{equation}
\Delta_{R} =  C_\delta \max_{f} (2 | \mathbf{n}_{R} \times \mathbf{r}_{f} |)
\end{equation}

In cells where the velocity gradient tensor has three real eigenvalues, no vortex axis is identified: the real eigenvector and the Liutex vector are set to zero, so that $\mathbf{n}_R=\mathbf{0}$ and $\Delta_R=0$. The outer length scale in these cells is then given by the lower bound $C_{lim}V^{1/3}$ of equation~\eqref{eq:cdes_amd_lim}. This fallback treatment directly affects the subgrid viscosity in the non-vortical regions of the initial shear layer, as discussed in \S\,\ref{sec:Perspective from vortex-axis identification}.

Proposal 4 is based on the eigenvector of the $S^2 + \Omega^2$ tensor, aligning with the $\lambda_2$ criterion \citep{jeong1995identification}. Since the tensor $M = S^2 + \Omega^2$ is inherently real and symmetric, its three eigenvalues are guaranteed to be real, with three mutually orthogonal eigenvectors. The local rotation axis, $\mathbf{n}_{\lambda_2}$, is defined by the eigenvector associated with the largest eigenvalue. This approach is numerically robust as it involves a real symmetric matrix, ensuring that a rotation axis is well-defined throughout the flow field without the need for fallback cases.

\begin{equation}
\Delta_{\lambda_2} =  C_\delta \max_{f} (2| \mathbf{n}_{\lambda_2} \times \mathbf{r}_{f} |)
\end{equation}

\begin{equation}
\nabla \mathbf{u} = \frac{\partial u_i}{\partial x_j} =
\begin{bmatrix}
\dfrac{\partial u}{\partial x} & \dfrac{\partial u}{\partial y} & \dfrac{\partial u}{\partial z} \\[10pt]
\dfrac{\partial v}{\partial x} & \dfrac{\partial v}{\partial y} & \dfrac{\partial v}{\partial z} \\[10pt]
\dfrac{\partial w}{\partial x} & \dfrac{\partial w}{\partial y} & \dfrac{\partial w}{\partial z}
\end{bmatrix}
\label{eq:velocity_gradient_app}
\end{equation}

\begingroup\footnotesize\setlength{\arraycolsep}{3pt}\medmuskip=2mu
\begin{equation}
S^2 = S_{ik}S_{kj} =
\begin{bmatrix}
S_{11}S_{11}+S_{12}S_{21}+S_{13}S_{31} & S_{11}S_{12}+S_{12}S_{22}+S_{13}S_{32} & S_{11}S_{13}+S_{12}S_{23}+S_{13}S_{33} \\[6pt]
S_{21}S_{11}+S_{22}S_{21}+S_{23}S_{31} & S_{21}S_{12}+S_{22}S_{22}+S_{23}S_{32} & S_{21}S_{13}+S_{22}S_{23}+S_{23}S_{33} \\[6pt]
S_{31}S_{11}+S_{32}S_{21}+S_{33}S_{31} & S_{31}S_{12}+S_{32}S_{22}+S_{33}S_{32} & S_{31}S_{13}+S_{32}S_{23}+S_{33}S_{33}
\end{bmatrix}
\label{eq:S2_app}
\end{equation}
\endgroup

\begingroup\footnotesize\setlength{\arraycolsep}{3pt}\medmuskip=2mu
\begin{equation}
\begin{aligned}
\Omega^2 &= \Omega_{ik}\Omega_{kj} \\
&= \begin{bmatrix}
\Omega_{11}\Omega_{11}+\Omega_{12}\Omega_{21}+\Omega_{13}\Omega_{31} & \Omega_{11}\Omega_{12}+\Omega_{12}\Omega_{22}+\Omega_{13}\Omega_{32} & \Omega_{11}\Omega_{13}+\Omega_{12}\Omega_{23}+\Omega_{13}\Omega_{33} \\[6pt]
\Omega_{21}\Omega_{11}+\Omega_{22}\Omega_{21}+\Omega_{23}\Omega_{31} & \Omega_{21}\Omega_{12}+\Omega_{22}\Omega_{22}+\Omega_{23}\Omega_{32} & \Omega_{21}\Omega_{13}+\Omega_{22}\Omega_{23}+\Omega_{23}\Omega_{33} \\[6pt]
\Omega_{31}\Omega_{11}+\Omega_{32}\Omega_{21}+\Omega_{33}\Omega_{31} & \Omega_{31}\Omega_{12}+\Omega_{32}\Omega_{22}+\Omega_{33}\Omega_{32} & \Omega_{31}\Omega_{13}+\Omega_{32}\Omega_{23}+\Omega_{33}\Omega_{33}
\end{bmatrix}
\end{aligned}
\label{eq:Omega2_full_app}
\end{equation}
\endgroup

\begin{equation}
\Omega^2 = \Omega_{ik}\Omega_{kj} =
\begin{bmatrix}
\Omega_{12}\Omega_{21}+\Omega_{13}\Omega_{31} & \Omega_{13}\Omega_{32} & \Omega_{12}\Omega_{23} \\[10pt]
\Omega_{23}\Omega_{31} & \Omega_{21}\Omega_{12}+\Omega_{23}\Omega_{32} & \Omega_{21}\Omega_{13} \\[10pt]
\Omega_{32}\Omega_{21} & \Omega_{31}\Omega_{12} & \Omega_{31}\Omega_{13}+\Omega_{32}\Omega_{23}
\end{bmatrix}
\label{eq:Omega2_reduced_app}
\end{equation}

Given that $S_{ij}$ is symmetric and $\Omega_{ij}$ is antisymmetric, $M_{ij}=S^2+\Omega^2$ is symmetric, with $M_{12}=M_{21}$, $M_{13}=M_{31}$ and $M_{23}=M_{32}$. Its six independent components are
\begin{equation}
\begin{aligned}
M_{11} &= S_{11}^2+S_{12}^2+S_{13}^2-\Omega_{12}^2-\Omega_{13}^2, \\
M_{22} &= S_{21}^2+S_{22}^2+S_{23}^2-\Omega_{21}^2-\Omega_{23}^2, \\
M_{33} &= S_{31}^2+S_{32}^2+S_{33}^2-\Omega_{31}^2-\Omega_{32}^2, \\
M_{12} = M_{21} &= S_{11}S_{12}+S_{12}S_{22}+S_{13}S_{32}+\Omega_{13}\Omega_{32}, \\
M_{13} = M_{31} &= S_{11}S_{13}+S_{12}S_{23}+S_{13}S_{33}+\Omega_{12}\Omega_{23}, \\
M_{23} = M_{32} &= S_{21}S_{13}+S_{22}S_{23}+S_{23}S_{33}+\Omega_{21}\Omega_{13}.
\end{aligned}
\label{eq:M_ij_app}
\end{equation}

\bibliographystyle{jfm}
\bibliography{refs}

\end{document}